\documentstyle[aps,prc,epsfig]{revtex}
\newcommand{\beq}{\begin{equation}}
\newcommand{\eeq}{\end{equation}}
\newcommand{\bea}{\begin{eqnarray}}
\newcommand{\eea}{\end{eqnarray}}
\newcommand{\bef}{\begin{figure}}
\newcommand{\eef}{\end{figure}}
\newcommand{\bce}{\begin{center}}
\newcommand{\ece}{\end{center}}
\newcommand{\eg}{{\it e.g.}}
\newcommand{\ie}{{\it i.e.}}
\newcommand{\etal}{{\it et al.}}
\def\lsim{\mathrel{\rlap{\lower4pt\hbox{\hskip1pt$\sim$}}
    \raise1pt\hbox{$<$}}}         
\def\gsim{\mathrel{\rlap{\lower4pt\hbox{\hskip1pt$\sim$}}
    \raise1pt\hbox{$>$}}}         
\begin{document}
\preprint{SUNY-NTG-00-41}
\tightenlines
\title
{Signatures of Thermal Dilepton Radiation at RHIC}
 
\author
{Ralf Rapp}
 
\address
{Department of Physics and Astronomy, State University of New York, 
    Stony Brook, NY 11794-3800, U.S.A.}

\date{\today} 

\maketitle
 
\begin{abstract}
The properties of thermal dilepton production from heavy-ion collisions 
in the RHIC energy regime
are evaluated for invariant masses ranging from 0.5 to 3~GeV. 
Using an expanding thermal fireball to model the evolution through 
both quark-gluon and hadronic phases various features of the spectra
are addressed. In the low-mass region, due to an expected large 
background, the focus is on possible medium modifications of the 
narrow resonance structures from $\omega$ and $\phi$ mesons, whereas in 
the intermediate-mass region the old idea of identifying QGP radiation
is reiterated including effects of chemical under-saturation 
in the early stages of central Au+Au collisions. 
\end{abstract}


\section{Introduction}
The identification of experimental signals from potential phase 
transitions in strong interaction matter associated with deconfinement 
and/or chiral symmetry restoration is the major goal of the (ultra-) 
relativistic heavy-ion program. 
Convincing evidence is likely to be achieved through a consistent 
combination of various suggested signatures. 
Among these is the copious production of thermal lepton pairs as 
originating from a supposedly thermalized fireball. Their relevance
comes in two facets depending on invariant mass.  

In the low-mass region (LMR, below $\sim$1~GeV) thermal radiation 
is governed by (in-medium) decays of the light vector mesons $\rho$, 
$\omega$ and $\phi$. With the dynamics in the light quark sector 
($u$, $d$, $s$) being driven by chiral symmetry and its spontaneous 
breaking in vacuum, medium modifications of the light vector mesons
should encode information on (the approach to) chiral restoration.
The challenge is then to discriminate the thermal contribution
from the numerous 'background' mostly consisting of meson decays after
freezeout of the interacting (hadronic) system.    

At sufficiently high masses (above 3~GeV or so), 
primordial processes such as Drell-Yan 
annihilation (characterized by a much harder slope than typical 
thermal sources) or  correlated decays of open anti/-charm (bottom) pairs 
are expected to prevail the spectra. Thus, the focus is on the  
narrow peaks from heavy vector
mesons $J/\Psi$, $\Upsilon$, $\dots$, in connection with deconfinement
signatures through their early dissolution. 
Descending from high masses 
thermal radiation is believed to become competitive below the 
$J/\Psi$ resonance. In the so-called  
intermediate-mass region (IMR), extending down to the $\phi$, 
the emission is continuum-like with a magnitude approximately 
given by perturbative $q\bar q$ annihilation. 
An enhanced dilepton yield in the IMR 
has been among the earliest suggestions~\cite{Sh80}
for a signal of QGP formation. The idea resides on the fact that 
at masses substantially larger
than typical temperatures, thermal occupation factors
strongly favor contributions from the hottest (early) stages of central
collisions where deconfined, perturbative matter ought to be formed.  

Dilepton measurements performed within the heavy-ion program
at the CERN-SpS over the last decade have demonstrated that the 
above issues can be thoroughly addressed. 
For Sulfur and Lead projectiles both the 
LMR~\cite{CERES,HELIOS3} and IMR~\cite{NA38-50,HELIOS3} clearly 
exhibit appreciable signals above the aforementioned 'background' 
sources, \ie, final state hadron decays as well as Drell-Yan and 
open charm contributions, respectively.
At the same time extensive theoretical analyses have shown that 
the experimental findings can be consistently understood in terms of 
thermal emission (coupled with strong medium effects in the LMR, 
see, \eg,  ref.~\cite{RW00} for a recent review),
indicating that one is indeed producing QCD matter in the vicinity
of the conjectured phase boundary. Nevertheless, the major part of the 
enhancement is still associated with the hot and dense hadronic 
phases in the course of the collisions. 

At the now operational collider experiments at RHIC, however, the plasma 
phase is anticipated to occupy much larger space-time volumes together
with higher initial temperatures. At the same time, primordial and final
hadron-decay sources increase.  
For the IMR various interplays have been elaborated in 
refs.~\cite{KMS92,SX93,Kaem95,GMRV96,MWCW96,Shu97,LVW98,LR98,GKP98,LK00}, 
and for the LMR assessments have been made, \eg, in ref.~\cite{qm99}.
The purpose of the present article is to provide an overall picture as 
well as an improved estimate of the thermal component in both the IMR and 
LMR, thereby utilizing our understanding as inferred from 
the investigations performed in the context of the SpS results.
 
The article is organized as follows. 
In Sect.~\ref{sec_corr} we will recall basic ingredients for the 
description of equilibrium  dilepton production rates from threshold 
to about 3~GeV. The emphasis will be on in-medium effects in the LMR, 
which will be assessed in the spectral function approach  
within hot hadronic matter as appropriate for RHIC conditions.  
In Sect.~\ref{sec_spec} we apply
the emission rates using an expanding fireball model around midrapidity
to calculate dilepton invariant mass spectra, suitable for comparison
with future measurements of the PHENIX detector.  
In the LMR special attention is paid to the manifestation of 
medium modifications of the light vector-meson resonances. In the IMR
the predictions for thermal yields will be confronted 
with Drell-Yan pairs to evaluate their significance for identifying
signals from an equilibrated partonic system. 
In particular, the consequences of chemical under-saturation 
in an early formed 'Gluon-Quark' Plasma will be assessed.  
In sect.~\ref{sec_concl} we summarize and conclude.

\section{Electromagnetic Correlation Function and Dilepton Production Rates} 
\label{sec_corr}

\subsection{Definition and Spectral Decomposition}
The general object for studying thermal dilepton emission is the
electromagnetic (e.m.) current-current correlator~\cite{Fe76,MT85}. 
In vacuum it is defined by the expectation value of 
the time-ordered product of e.m. currents, 
\beq
\Pi_{\rm em}^{\circ,\mu\nu}(q)=-i\int d^4x \ e^{iqx} \ 
\langle 0|{\cal T} j_{em}(x) j_{em}(0)|0\rangle  . 
\label{Piem}
\eeq
The correlation function provides a common framework not only for 
different hadronic approaches but also for quark-gluon based calculations. 
This feature is particularly evident in free space where the 
following decomposition of its imaginary part can be given:
\begin{eqnarray}
{\rm Im} \Pi_{\rm em}^{\mu\nu}  = 
\left\{ \begin{array}{ll}
 \sum\limits_{V=\rho,\omega,\phi} ((m_V^{(0)})^2/g_V)^2 \
{\rm Im} D_V^{\mu\nu}(M,\vec q;\mu_B,T) & , \ M \le M_{dual}
\vspace{0.3cm}
\\
(-g^{\mu\nu}+q^\mu q^\nu/M^2) \ (M^2/12\pi) \ N_c
\sum\limits_{q=u,d,s} (e_q)^2  & , \ M \ge M_{dual} \ .
\end{array}  \right.
\label{ImPiem}
\end{eqnarray}
From total $e^+e^-\to hadrons$ cross section data it is well 
established that the low-mass part of the correlator can be accurately 
described by 
the light vector mesons within the Vector Dominance Model (VDM),
whereas above a 'duality threshold'
of about $M_{dual}\simeq 1.5$~GeV free quark-antiquark states 
satisfactorily saturate the strength (within 20\% or so).   

The thermal dilepton production rate per unit 4-volume and 
4-momentum is directly related to the (spin-averaged) imaginary part 
of the in-medium (retarded) e.m. correlator through 
\bea
\frac{dR_{l^+l^-}}{d^4q} &=& - 2 \ f^B(q_0;T) \ L_{\mu\nu} \ 
                           {\rm Im} \Pi_{\rm em}^{\mu\nu} 
\nonumber\\
     &=& -\frac{\alpha^2}{\pi^3 M^2} \ f^B(q_0;T) \ 
                {\rm Im} \Pi_{\rm em}(q_0,\vec q;T,\mu_B) \ , 
\eea  
where the lepton tensor $L_{\mu\nu}$ encodes the propagation of a virtual 
photon and its subsequent decay into a pair of leptons.

\subsection{Medium Effects in the Low-Mass Region (LMR)}
In the LMR medium effects arise from changes of the   
vector-meson spectral functions in hot hadronic matter. At RHIC energies, 
due to the low stopping of the interpenetrating nuclei, one would 
not expect baryons to play a significant role at central rapidities.
However, even for a vanishing net baryon density this may not 
necessarily be true once accounting for 
thermal excitations of $N\bar N$ pairs in connection with the  
rather strong meson-baryon interactions (as compared to the relatively 
weaker meson-induced interactions).
In the following we will employ standard many-body techniques  
to implement (hadronic) selfenergy corrections into the propagation 
of the three light vector mesons $\rho$, $\omega$ and $\phi$.

\subsubsection{$\rho$-Meson}
For the $\rho$-meson propagator,
\beq
D_\rho(M,q;T)=\left[M^2-{m_\rho^{(0)}}^2-\Sigma_{\rho\pi\pi}
-\Sigma_{\rho M} -\Sigma_{\rho B} \right]^{-1}
\eeq
an extensive literature about hadronic in-medium modifications
has emerged over the last 10 years (see \cite{RW00} for a recent 
review). 
Here, we incorporate finite temperature effects along the lines of 
ref.~\cite{RG99}, where the 
total in-medium selfenergy has been obtained in terms of 
polarizations of its pion cloud as well as 
resonant interactions with surrounding pions, kaons and rho mesons, 
$M=\pi,K,\bar K, \rho$. In addition, modifications due  
to nucleons as well as the most abundant baryonic resonances
($B=N, \Delta(1232), N(1440), N(1520)$) are included 
following refs.~\cite{RCW97,RW99}. 
Since the $\rho^0$-meson is an eigenstate under antiparticle conjugation,
it equally interacts with antibaryons, thereby 
effectively doubling the impact  of the baryonic component for 
$\mu_B$ close to zero. {\it E.g.},  at $(T,\mu_B)=(180,0)$~MeV one has 
$\varrho_B=\varrho_{\bar B}\simeq 0.46\varrho_0$ which is appreciable 
(including all nonstrange resonances up to $m_B=1.7$~GeV as well as the 
lightest hyperons; $\varrho_0=0.16$~fm$^{-3}$ denotes ground state nuclear
matter density). 
 

\subsubsection{$\omega-Meson$}
The same approach as for the $\rho$ is now adopted  
for the $\omega$ meson 
(see also refs.~\cite{Sh91,Ha95,Alam99} for somewhat
different treatments). First, its vacuum selfenergy is constructed 
from a combination of the direct $\omega\to \rho\pi$
and the $\omega\to 3\pi$ decays derived from the interaction 
vertices (isospin structures suppressed)~\cite{Sa62,GSW62}
\bea
{\cal L}_{\omega\rho\pi}&=& G_{\omega\rho\pi} \ \epsilon_{\mu\nu\sigma\tau} \ 
  \omega^\mu \ \partial^\nu \rho^\sigma \ \partial^\tau \pi 
\\
{\cal L}_{\omega 3\pi}  &=& G_{\omega 3\pi} \ \epsilon_{\mu\nu\sigma\tau}
 \ \omega^\mu \ \partial^\nu \pi \ \partial^\sigma \pi \ \partial^\tau \pi \ , 
\eea
respectively. $G_{\rho\pi\omega}$ is taken from ref.~\cite{RG99}
which (together with a hadronic formfactor) gives the correct radiative
decay width $\Gamma_{\omega\to\pi\gamma}$ (using VDM) 
and about 50\% of the full hadronic width. $G_{3\pi}$ is then 
adjusted to reproduce the missing part.  
Medium effects in the $\pi\rho$ cloud are introduced through thermal 
Bose enhancement factors
(evaluated in the Matsubara framework) as well as 
the in-medium $\rho$ spectral function from above, \ie, 
\bea
\Gamma_{\omega\to\rho\pi}(s) &=& \frac{2~G_{\omega\rho\pi}^2}{8\pi} \ 
\ \int\limits_0^{M_{max}} \frac{M dM}{\pi} \ A_\rho(M) \  
[1+f^\pi(\omega_\pi)+f^\rho(\omega_\rho)] \ 2 q_{cm}^3 \  
F_{\omega\rho\pi}(q_{cm})^2  
\label{Gamorp}
\\
\Gamma_{\omega\to3\pi}(s) &=& \frac{9~G_{\omega 3\pi}^2}{192\pi^3} 
\int\limits_{m_\pi}^{\omega_{1,max}} d\omega_1 
\int\limits_{m_\pi}^{\omega_{2,max}} d\omega_2 \ 
\left[1+ \lbrace\sum\limits_{i=0}^2 f^\pi(\omega_i)\rbrace+
f^\pi(\omega_0) \lbrace f^\pi(\omega_1) +f^\pi(\omega_2) \rbrace +
f^\pi(\omega_1) f^\pi(\omega_2) \right] 
\nonumber\\
&  & \qquad \qquad \qquad \qquad \qquad \times 
\left[p_1^2 p_2^2 -\frac{1}{4} (p_0^2-p_1^2-p_2^2)\right] \
F_{\omega 3\pi}(p_0,p_1,p_2)^2    
\label{Gom3p}
\eea
with $M_{max}=\sqrt{s}-m_\pi$,  $\omega_{1,max}=\sqrt{s}-2m_\pi$ and 
$\omega_{2,max}=\sqrt{s}-(\omega_1+m_\pi)$ (dynamical modifications 
of the pion propagators are smaller and have been neglected).  
The $\omega$ can also be absorbed through the inelastic reaction 
$\omega\pi\to \pi\pi$. The corresponding collisional broadening
has been calculated in ref.~\cite{Ha95} which we include via 
\beq
\Gamma_{\omega\pi\to\pi\pi} \simeq (7{\rm MeV}) \ 
\left(\frac{T}{T_0}\right)^3 \   
\eeq
($T_0=150$~MeV). 
We furthermore compute an $\omega$-selfenergy from scattering off 
thermal pions inducing $b_1(1235)$ resonance formation~\cite{Sh91,Ha95}, 
which has the largest known coupling to $\omega\pi$ states. Using the 
interaction vertex 
\beq
{\cal L}_{\omega\pi b_1} = G_{\omega\pi b_1} \ b^\mu \ 
(g_{\mu\nu} p\cdot q -q_\mu p_\nu ) \ \omega^\nu \ \pi \ ,   
\eeq     
the selfenergy takes the form
\beq
\Sigma_{\omega\pi b_1}(M,q)=G_{\omega\pi b_1}^2 
\int \frac{p^2 dp dx}{(2\pi)^2 2\omega_\pi(p)} \ [f^\pi(\omega_\pi(p))-
f^{b_1}(q_0+\omega_\pi(p)] \ F_{\omega\pi b_1}(q_cm)^2 \ D_{b_1}(s) 
 \ v_{\omega\pi b_1}(p,q)  \  ,
\eeq
where the vertex function $v_{\omega\pi b_1}(p,q)$ and the hadronic 
(dipole) formfactor $F_{\omega\pi b_1}$ are in close analogy  
to the $\rho\pi a_1$ case~\cite{RG99}. The coupling constant and cutoff
parameter are fitted to the empirical hadronic and 
radiative decay widths $\Gamma_{b_1 \to \pi\omega,\pi\gamma}$.   

In nuclear matter, according to the coupled channel analysis for vector 
meson-nucleon scattering of ref. \cite{FLW00}, the dominant coupling of the 
$\omega$  proceeds through resonant $N(1520)N^{-1}$ excitations (with a 
strength similar to that for the $\rho$ meson), as well as 
$N(1650)N^{-1}$ ones.  
Modifications induced in the pion cloud are presumably less 
significant~\cite{KKW97}, especially when 
using soft formfactors for the $\pi NN$ and $\pi N\Delta$ as suggested 
by the analysis of photoabsorption spectra on the nucleon and 
nuclei~\cite{RUBW}. In line with the arguments given above (see also
ref.~\cite{RW99}), we evaluate the $N(1520)N^{-1}$ and $N(1650)N^{-1}$ 
loops at an effective nucleon density 
\beq
\varrho_{eff}=\varrho_N+\varrho_{\bar N}+0.5~(\varrho_B+\varrho_{\bar B})
\label{rhoeff}
\eeq
to account for the effects of higher resonances as well as antibaryons. 
The in-medium $\omega$ propagator finally reads 
\beq
D_{\omega}=\left[ M^2-m_\omega^2+
i m_\omega (\Gamma_{3\pi}+\Gamma_{\rho\pi}+\Gamma_{\omega\pi\to\pi\pi})
-\Sigma_{\omega\pi b_1} - \Sigma_{\omega B} \right]^{-1} \ . 
\eeq

\subsubsection{$\phi$-Meson}  
For the $\phi$ meson various evaluations of medium effects based 
on hadronic rescatterings at finite temperature have indicated 
quite  moderate changes of both its mass and 
width~\cite{KS94,HG94,Ha95,AGPR97}. This also holds true for direct
$\phi N$ interactions as indicated by photonuclear production of $\phi$ 
mesons as well as the absence of baryonic resonances with 
explicitly measured $\phi N$ decay channels\footnote{This behavior is quite 
contrary to the light quark states $\rho$ and $\omega$. The reason for 
that is presumably related to the OZI rule: whereas in a $\rho$-/$\omega$-$N$
interaction the antiquark in the vector meson can easily annihilate and 
convert into excitation energy of a baryonic resonances, the equivalent 
mechanism  for the antistrange quark in a $\phi$-$N$ interaction is 
strongly suppressed.}.    
Therefore we do not 
attempt a microscopic description of this part of the  $\phi$ selfenergy 
but merely parameterize the finite-temperature collisional broadening 
guided by the results of ref.~\cite{Ha95} as  
\beq
{\rm Im}~\Sigma_\phi^{coll}(T)= -m_\phi \ \Gamma_\phi^{coll}(T_0) \ 
\left(\frac{T}{T_0}\right)^3 
\eeq 
with $T_0=150$~MeV, $\Gamma_\phi^{coll}(T_0)=9$~MeV.

In addition, there can arise width modifications from the dressing
of the anti-/kaon cloud through in-medium $\phi\to K\bar K$ decays.  
Here even small effects in the kaon propagators can have appreciable 
impact on the $\phi$ spectral function due to the proximity of 
the $\phi$ mass to the two-kaon threshold (the $p$-wave nature of 
the $\phi\to K\bar K$ decay further enhances this sensitivity).   
We assess this by evaluating the  
$\phi\to K\bar K$ width in terms of in-medium modified (anti-)
kaon spectral functions, \ie, 
\bea
{\rm Im}~\Sigma_{\phi K\bar K}(M;T) &=& -m_\phi \ 
\Gamma^{vac}_{\phi\to K\bar K}  
\int\limits_{0}^{M} \ \frac{m_1 dm_1}{\pi} A_K(m_1) 
\nonumber\\ 
 & & \quad \times\int\limits_0^{M-m_1} \frac{m_2 dm_2}{\pi} \ A_{\bar K}(m_2) 
\left(\frac{p_K}{p_K^{on}}\right)^3 \ F_{\phi K\bar K}(p_K)^2 \
[1+f^K(\omega_1)+f^{\bar K}(\omega_2)] \    
\eea 
($\Gamma^{vac}_{\phi\to K\bar K}=3.7$~MeV: on-shell width in free space),
and again neglect possible modifications of the real part. The (anti-) 
kaon spectral functions are given by
\beq 
A_{K,\bar K}(m)=-2~{\rm Im} D_{K,\bar K}(m)=
-2~{\rm Im} \frac{1}{m^2-m_K^2-\Sigma_{K,\bar K} } \ . 
\eeq  
For the finite-temperature part of $\Sigma_{K,\bar K}$, which does 
not distinguish between kaons and antikaons, we approximate the results of 
ref.~\cite{Sh91} as 
\bea
{\rm Im} \Sigma_{K,\bar K}(T) &\simeq&
 -m_K \ \Gamma_{K,\bar K}(T) = -m_K \ (20{\rm MeV}) \ 
\left(\frac{T}{T_0}\right)^3 
\nonumber\\
{\rm Re} \Sigma_{K,\bar K}(T) &\simeq& -m_K \ (10{\rm MeV}) \ . 
\eea
In nuclear matter kaons and antikaons behave quite differently. 
The $KN$ interaction is rather weak and slightly repulsive. A corresponding
selfenergy has been extracted as~\cite{KSW95,OR00}
\beq
\Sigma_K(\varrho_N)\simeq 
m_K \  (0.13m_K) \ \left(\frac{\varrho_N}{\varrho_0}\right) \ .  
\eeq
The antikaon, on the other hand, couples to a number of hyperon excitations.
A detailed recent coupled channel calculation~\cite{RO00} predicts  
an appreciable  $\bar K$ width of up to $100$~MeV already at nuclear 
saturation density, together with significant attraction as well. 
Accordingly, we employ
\beq
\Sigma_{\bar K}(m;\varrho_N)\simeq 
m_K \ \left(\frac{\varrho_N}{\varrho_0}\right) (-60,-i~90) {\rm MeV} \ . 
\eeq
Notice that the  situation reverses for the interaction with antinucleons, 
\ie, 
\bea
\Sigma_{K}(\varrho_{\bar N})=\Sigma_{\bar K}(\varrho_N) 
\\
\Sigma_{K}(\varrho_{N})=\Sigma_{\bar K}(\varrho_{\bar N})
\eea
Thus, for central rapidities at RHIC, where the {\em net} baryon density
in the hadronic phase is small, the combined contributions
to the real part of both the anti-/kaon selfenergies practically vanish.
This is not the case for the imaginary parts which will be explicitly 
included.  

Finally, we take into account the 13\% branching ratio of the free 
$\phi$ decay into $\pi\rho$, and treat medium modifications thereof 
in analogy to the $\omega\to\rho\pi$ decay, eq.~(\ref{Gamorp}). 

Combining all contributions yields an in-medium $\phi$ propagator 
of the form
\beq
D_{\phi}=\left[ M^2-m_\phi^2-\Sigma_{\phi K\bar K}-
\Sigma_{\phi\rho\pi} -\Sigma_\phi^{coll} \right]^{-1} \ .
\eeq

\subsubsection{Spectral Functions}
Fig.~\ref{fig_ImDV} summarizes the appearance of the light vector-meson 
spectral functions (=imaginary part of the propagators) 
that will be employed in the calculation of dilepton 
spectra below: both light quark states $\rho$ and $\omega$ undergo
a strong broadening towards the expected phase boundary (one also
observes a slight upward mass shift caused  by repulsive
parts in the selfenergy which are mainly due to baryonic particle-hole 
excitations~\cite{RW99,FLW00}, as well as medium effects in the pion tadpole
graph in case of the $\rho$~\cite{UBRW00}). The $\phi$
seems to retain more of its resonance structure, although, at the highest
temperature, the hadronic rescattering has increased its vacuum width 
by over a factor of 7 to $\sim$~32~MeV. 
\bef
\vspace{-0.3cm}
\bce
\epsfig{file=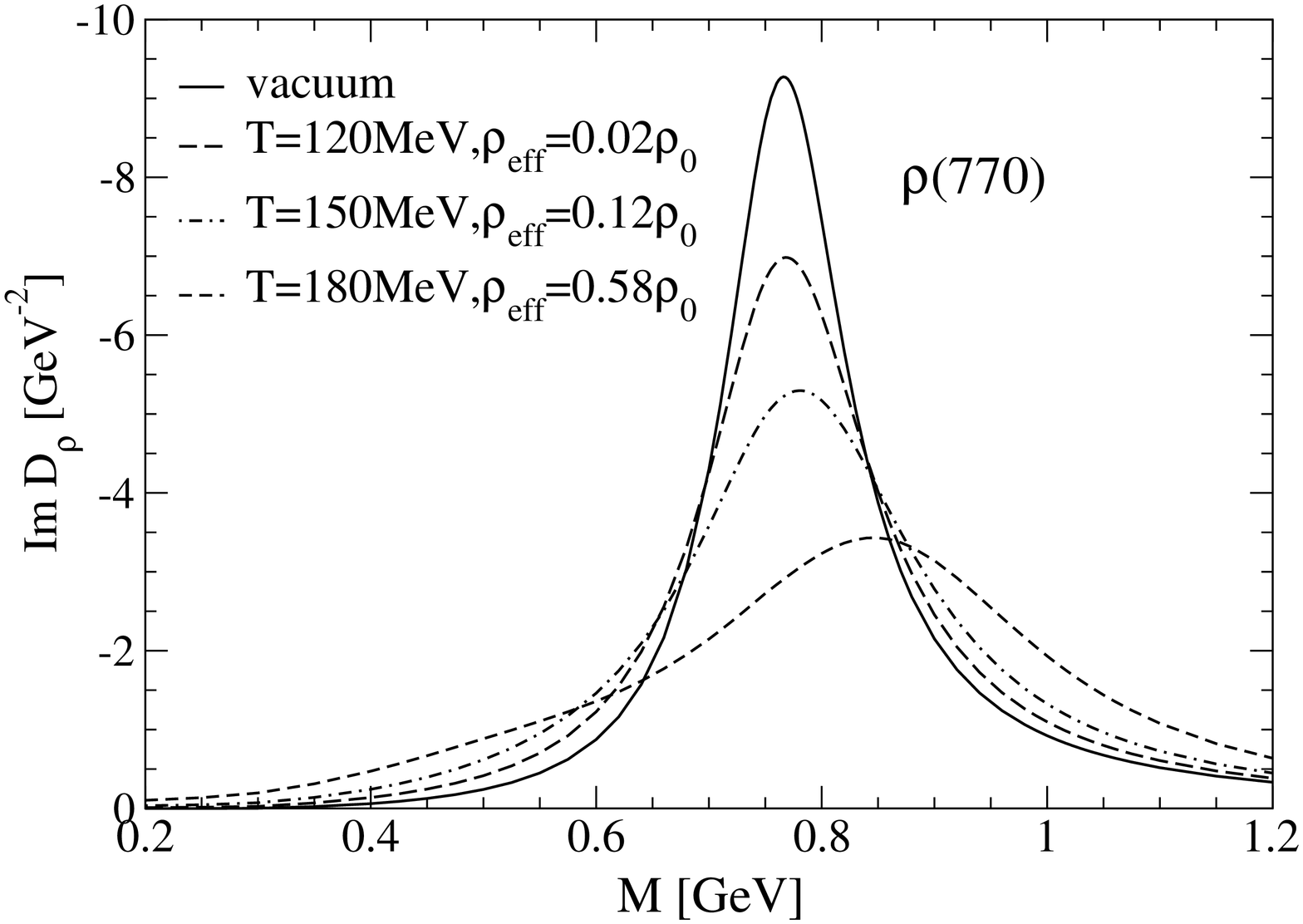,width=6.8cm,angle=-90}
\epsfig{file=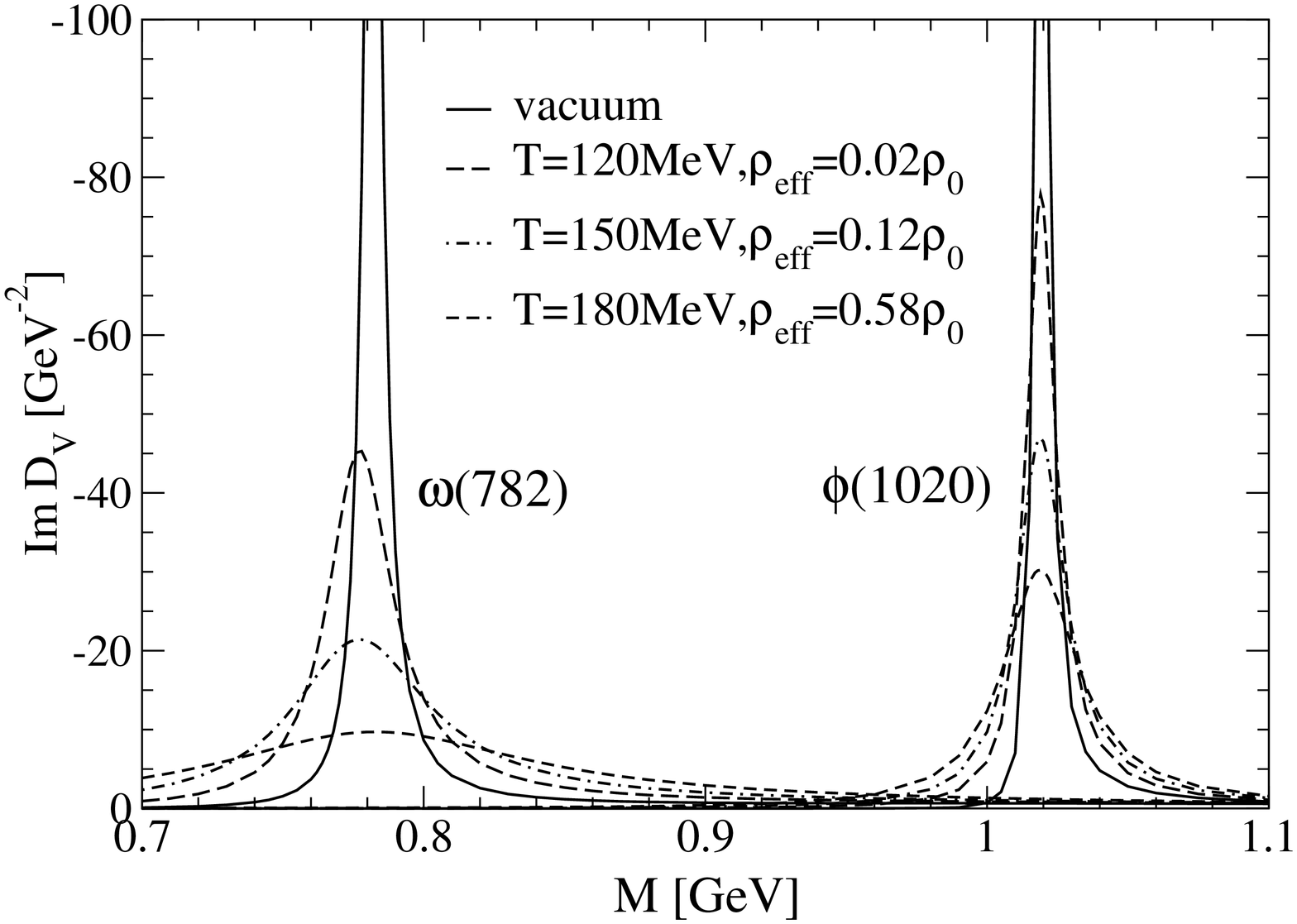,width=6.8cm,angle=-90}
\ece
\caption{Spectral functions of the light vector mesons $\rho$ (left
panel) and $\omega$, $\phi$ (right panel) in vacuum
(solid lines) as well as in hot baryon-poor hadronic matter as expected 
under RHIC conditions: $(T,\mu_N)=(120,91)$~MeV (long-dashed lines),
$(T,\mu_N)=(150,40)$~MeV (dashed-dotted lines) and 
$(T,\mu_N)=(180,27)$~MeV (short-dashed lines). For the definition
of $\varrho_{eff}$ see eq.~(\protect\ref{rhoeff}).} 
\label{fig_ImDV}
\eef
It is interesting to note that at $T=180$~MeV, 
where for the chosen value of $\mu_B=27$~MeV the {\em net} baryon density 
amounts to only $0.14\rho_0$, the combined effect of baryons and 
antibaryons causes an appreciable further broadening of both the 
$\rho$ and $\omega$ spectral functions as compared to the pure 
meson gas results, see fig.~\ref{fig_ImDV180}. 
This does not apply for the $\phi$ meson.
It would be very illuminating to scrutinize 
these features with high resolution/statistics experiments 
in more baryon-dominated systems
as created in fixed target heavy-ion collisions.  
\bef
\bce
\epsfig{file=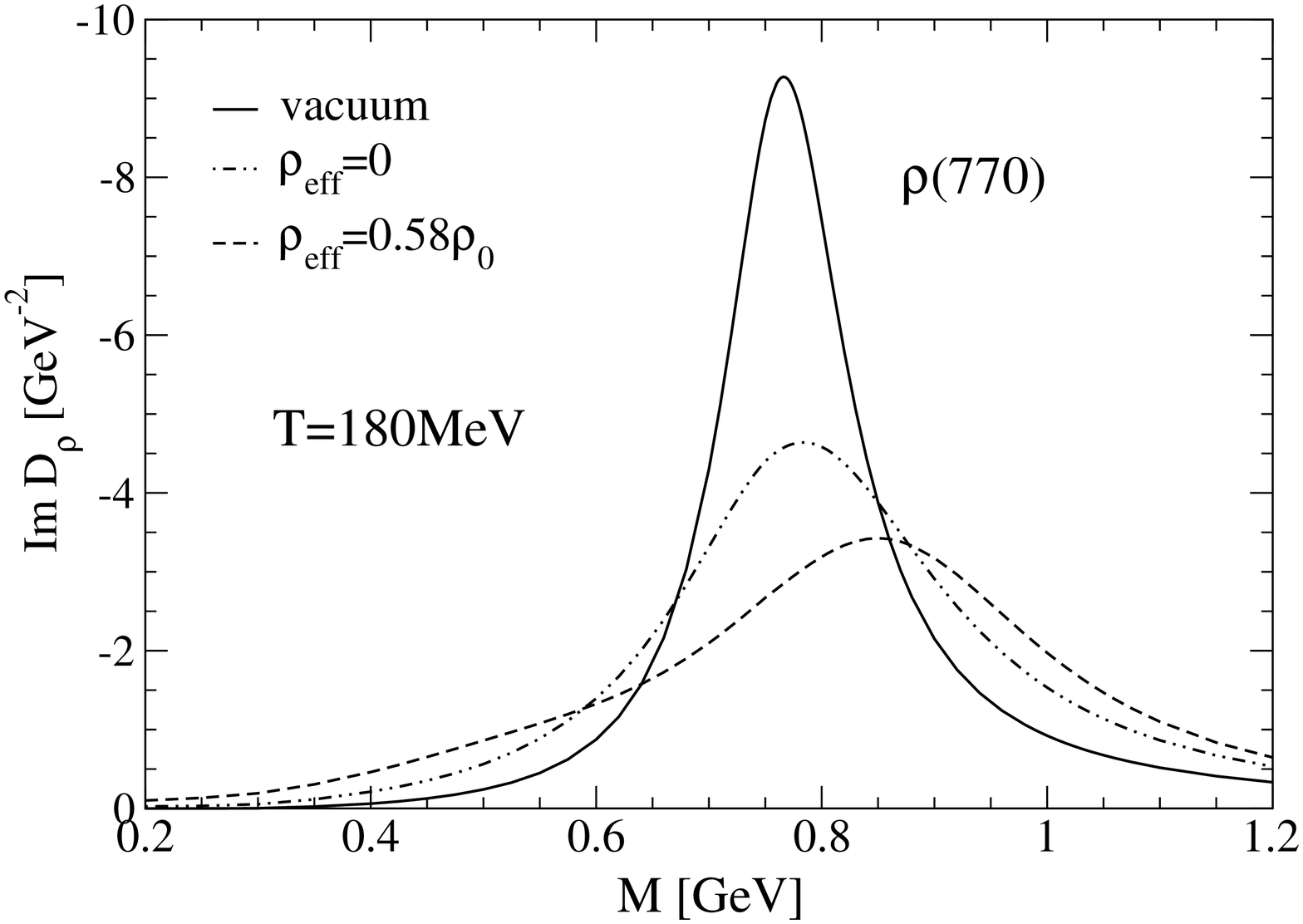,width=6.8cm,angle=-90}
\epsfig{file=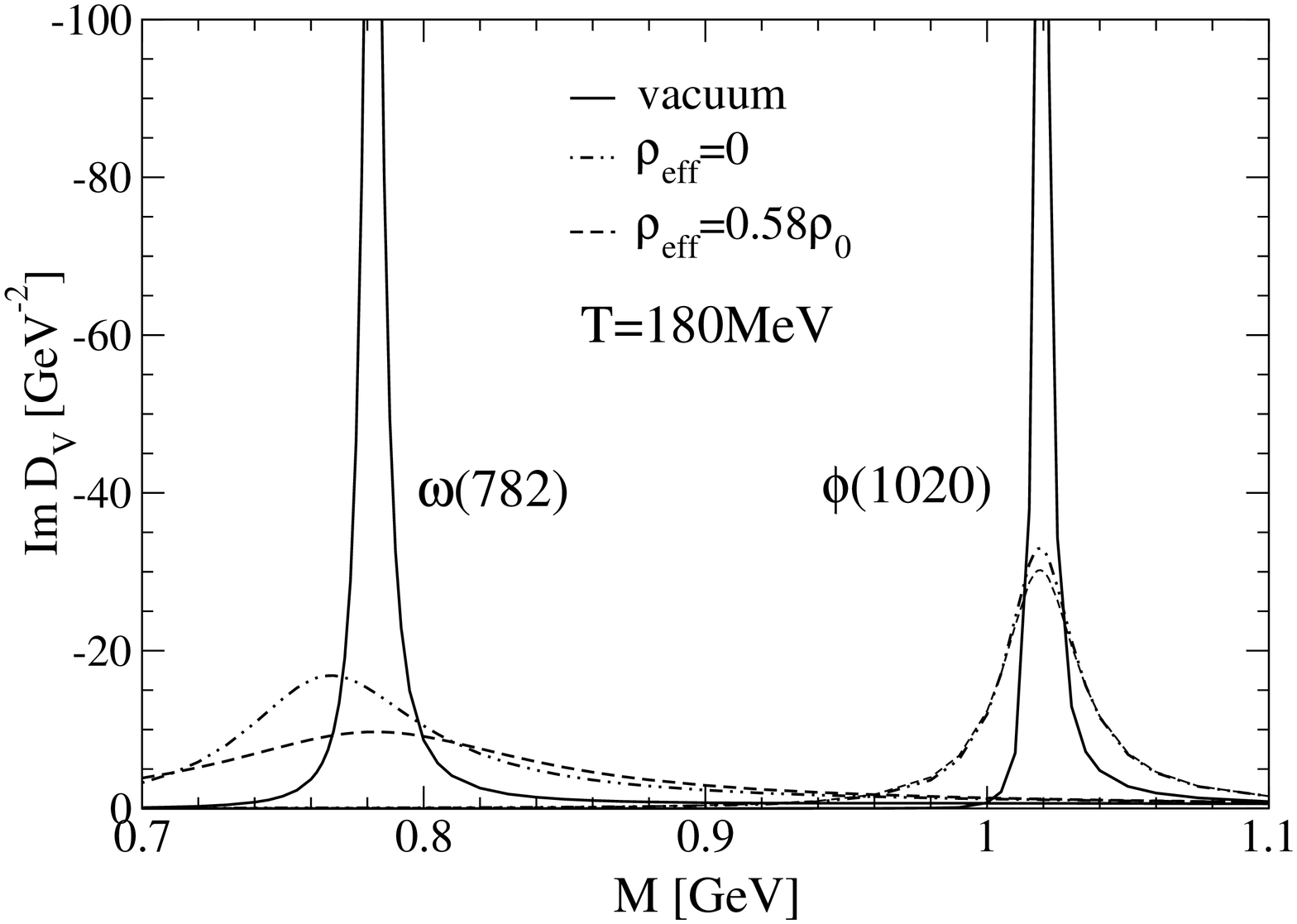,width=6.8cm,angle=-90}
\ece
\caption{$\rho$ (left panel) and $\omega$, $\phi$ (right panel) spectral
functions at $T=180$~MeV with (short-dashed lines, $\mu_B=27$~MeV) and 
without (dashed-double-dotted lines) the effects of anti-/baryons as 
expected for central rapidities at RHIC.} 
\label{fig_ImDV180}
\eef

\subsection{Intermediate Mass Region (IMR)}
The nonperturbative mechanisms related to the deconfinement/chiral
restoration transition necessarily imply substantial medium effects
for the light hadronic bound states, as discussed in the previous 
section. Beyond invariant masses of 
$M\simeq1.5$~GeV, however, the interactions in the free vector 
correlator essentially become perturbative, cf. eq.~(\ref{ImPiem}). 
Naively, one might anticipate in-medium corrections  to be of order 
$\alpha_s$ or ${\cal O}(T/M)$, 
${\cal O}(\mu_q/M)$, but theoretically the impact of 'soft' corrections 
to the finite-temperature $q\bar q$ rates is not very well under
control, see, \eg, refs.~\cite{BPW90,CW95,LWZH99,PT00}. 
For the (closely related) Drell-Yan annihilation process, corrections 
of similar nature are subsumed in the famous $K$-factors, 
which, for modern parton distribution functions (\eg, GRV-94~\cite{GRV95}), 
are around 1.3-1.5.  This should roughly 
reflect the uncertainty when employing the lowest-order thermal 
$q\bar q\to l^+l^-$ annihilation for the QGP phase 
as done below (and might be considered 
as a lower limit of the true emissivity). Making furthermore use 
of 'duality' arguments~\cite{LR95,RW00,RS00} we will in fact apply the 
corresponding dilepton production rate~\cite{CFR87}, 
\beq
\frac{d^8N_{\mu\mu}}{d^4xd^4q} = \frac{\alpha^2}{4\pi^4}
\frac{T}{q} f^B(q_0;T) \sum\limits_q e_q^2
\ln \frac{\left(x_-+y\right) \left( x_++\exp[-\mu_q/T]\right)}
{\left(x_++y\right) \left( x_-+\exp[-\mu_q/T]\right)} 
\label{dualrate}
\eeq 
also in the hadronic phase, extrapolating it down to masses of 
about 1.1~GeV.
The latter is motivated by the observation that lowest-order
in temperature effects~\cite{Dey90} already amount to a  
reduction of the 'duality threshold' towards the $\phi$ 
resonance (see, \eg, the discussion in \cite{Rapp99,RW00}).

\subsection{Thermal Rates}

In fig.~\ref{fig_rates} we compare the 3-momentum integrated 
dilepton production rates in thermal equilibrium, 
\beq
\frac{dR_{ee}}{dM^2}=\int \frac{d^3q}{2q_0} \frac{dR_{ee}}{d^4q} 
\eeq
at fixed temperature and density 
for 3 different ways of evaluating the electromagnetic correlator:  
using hadronic degrees of freedom with (solid line)
or without (dashed lines) medium effects in the vector-meson 
spectral functions, as well as perturbative quark-antiquark 
annihilation (dashed-dotted lines).   
\bef[!th]
\bce
\epsfig{file=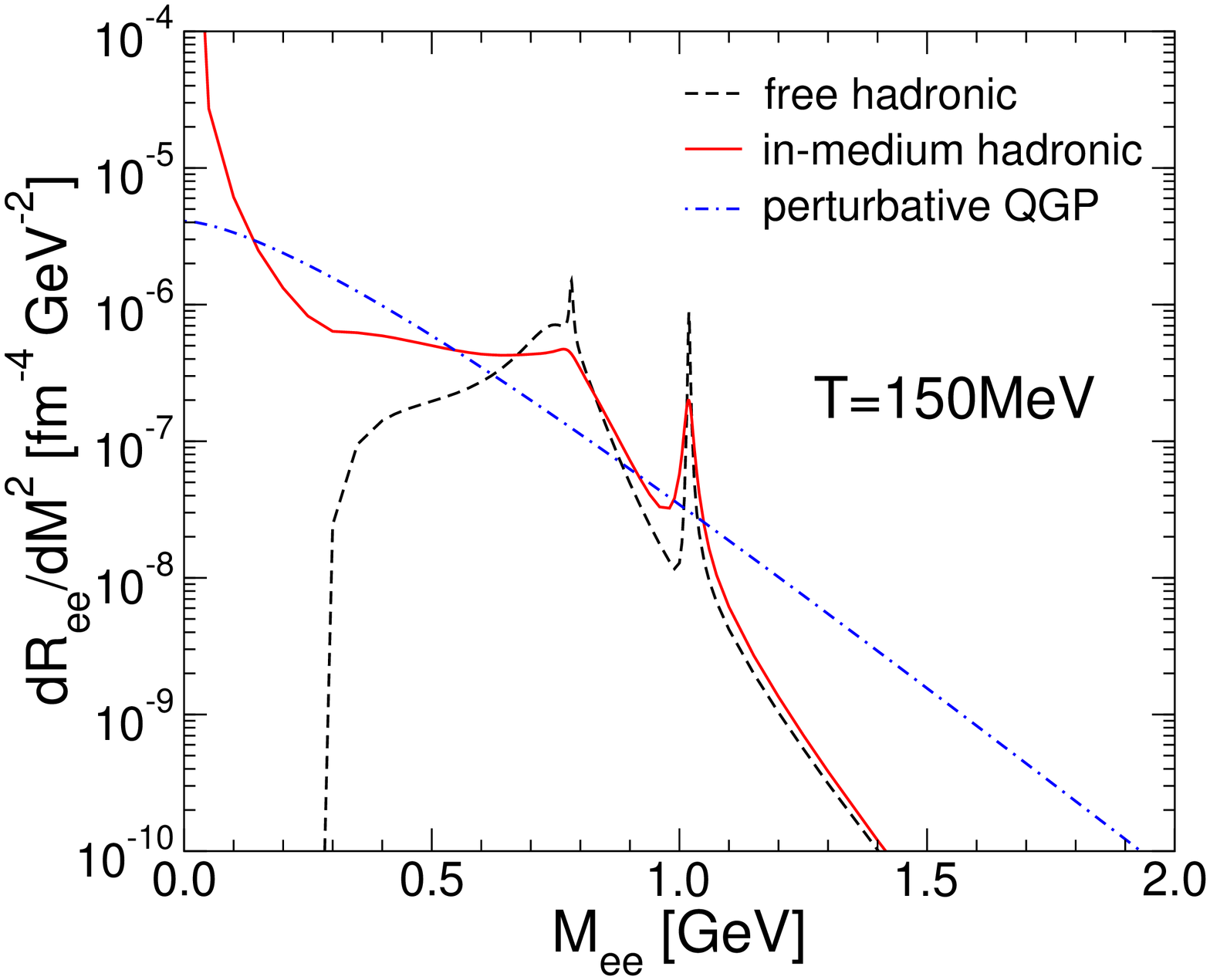,width=7.2cm,angle=-90}
\hspace{-1.0cm}
\epsfig{file=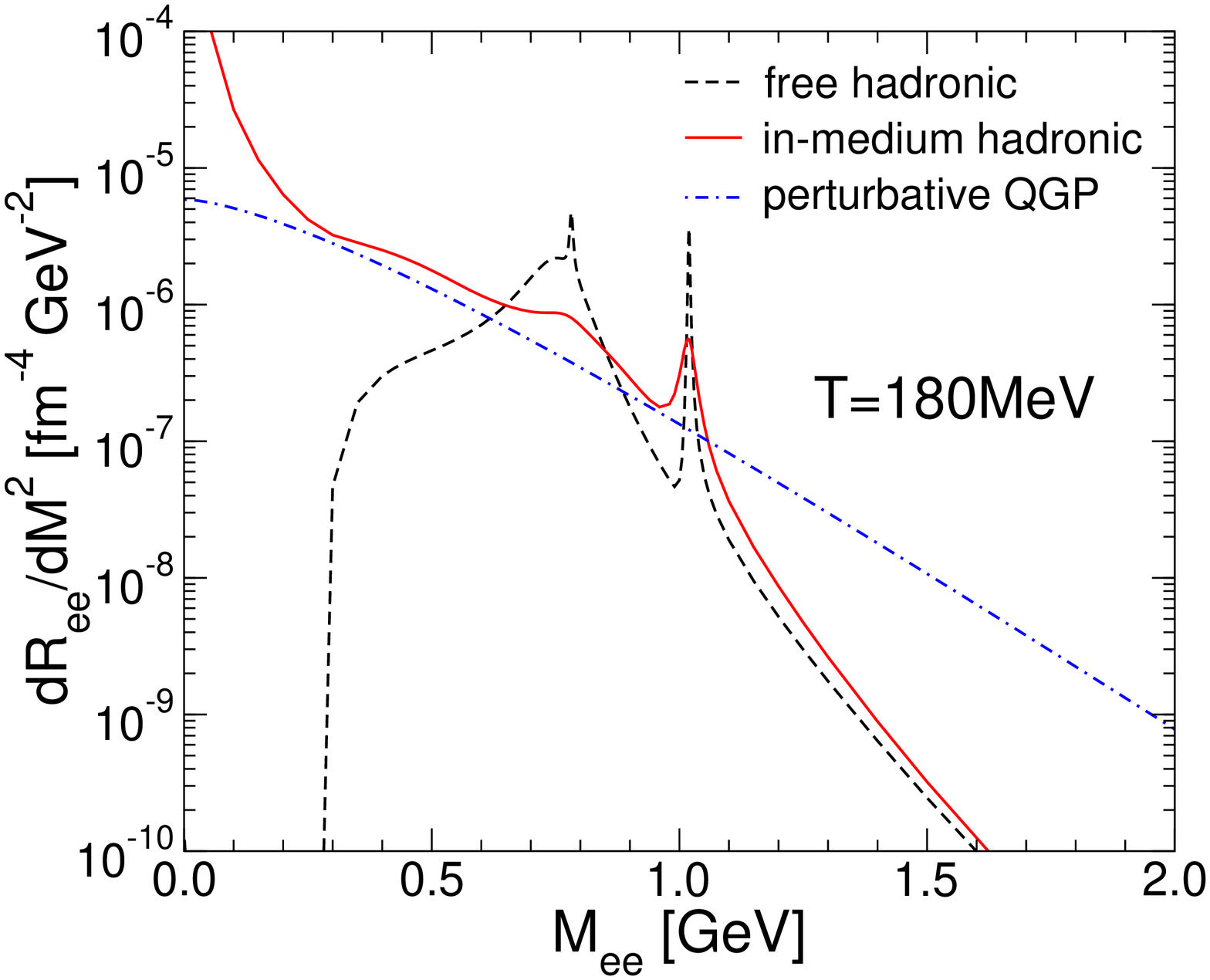,width=7.2cm,angle=-90}
\ece
\caption{3-momentum integrated  thermal dilepton production rates
as appropriate for the hadronic phase at RHIC (left panel:
$(T,\mu_N)=(150,60)$~MeV, right panel: $(T,\mu_N)=(180,27)$~MeV).
The dashed lines are obtained when using the free $\rho$, $\omega$
and $\phi$ spectral functions, whereas the full lines follow upon
inclusion of hadronic medium effects as described in the previous
section; the dashed-dotted is the result for lowest order $\alpha_s$
quark-antiquark annihilation.}
\label{fig_rates}
\eef
The hadronic calculations have built in contributions
from  2- and 3-pion as well as $K\bar K$ states through the $\rho$
and $\omega$ as well as the $\phi$, respectively. These
are sufficient to saturate the strength in the electromagnetic correlator
from the two-pion threshold up to about 1~GeV.
Beyond, 4-pion states start to become dominant, most notably through
$\pi a_1$ and $\pi \omega$ annihilation~\cite{LG98},
which are rather structureless and merge into the aforementioned
quark-hadron duality from $M_{dual}\ge 1.5$~GeV on (in vacuum).

The medium effects in the spectral functions are directly  
reflected in the dilepton rates. The thermal Bose 
factor in the latter augments the enhancement on the low-mass side of 
the resonances (a factor of 3-10 over the free case 
around  invariant masses $M\simeq 0.4$~GeV for $T=150-180$~MeV).  
To a large extent, the underlying processes can be associated with 
(in matter) Dalitz decays $\omega, a_1(1260) \to \pi e^+e^-$ or 
$N^*\to N e^+e^-$, being proportional to the imaginary parts of the 
vector-meson selfenergies. At the same time, a strong broadening of the
resonance structures is implied as a result of resumming the large 
imaginary parts~\cite{Rapp99,RG99} in the propagators.
Taken together, both features seem to drive the hadronic in-medium 
curve towards perturbative $q\bar{q}$ annihilation:  
at $T=180$~MeV the magnitude and slope of the former
in the $M\simeq~0.5$~GeV region are quite reminiscent to 
the plasma rate (not so at $T=150$~MeV); even in the $\omega$
resonance region, where the maximum of the free rate is suppressed
by a factor of $\sim$6, the in-medium hadronic and the $\bar qq$ result
agree within a factor of 2. On the other hand, at $T=150$~MeV 
where the in-medium suppression amounts to a factor of $\sim$3, 
in-medium hadronic and $\bar qq$ rates differ by a factor
of $\sim$4. 

One concludes that the indications for quark-hadron duality in dilepton 
production close to the phase boundary, as put forward in 
ref.~\cite{RW99} for SpS conditions, also (approximately) persist  
in the (almost) net baryon-free regime at RHIC.  

\subsection{'Background' Sources}
To be able to extract information on thermal dilepton radiation in the 
complicated environment of heavy-ion reactions
a reliable assessment of the multiple other sources is mandatory. 
At low masses these are mostly Dalitz decays of $\pi^0$,  $\eta$
and $\omega$ mesons as well as the direct decays of produced $\omega$'s
and $\phi$'s in the final state (commonly denoted as 'hadronic cocktail'). 
Restricting ourselves to invariant 
masses above 0.5~GeV, we explicitly compute the latter two 
contributions based on the $\omega$- and $\phi$-abundances as obtained 
from the thermal fireball model at thermal break-up, 
\beq
\frac{dN_{V\to ee}^{cktl}}{dM}= \frac{\alpha^2}{\pi^3 M} \ F_V(M)^2
\ V_{fo} \ \int d^3q \ f^V(q_0;T_{fo})  \ ,  
\label{cktl}
\eeq 
Here, $V_{fo}$ is the (final) fireball 3-volume, $f^V$ are Bose 
distributions and 
\beq
F_V(M)^2 = m_V^2 \ |D_V^\circ(M)|^2 
\eeq
the (free) electromagnetic formfactors ($V=\omega, \phi$). Note that in 
such a framework the contribution from $\rho$-mesons, due to their short 
lifetime ($\tau_\rho\le \tau_{\rho}^{vac}=1.3$~fm/c), is more 
appropriately evaluated through a slight extension
of the fireball evolution, since their medium modifications 
around freezeout will still be significant. The same argument holds for 
higher resonances in the $M\simeq 1.5$~GeV region ($a_1$, $\rho'$, etc.), 
which effectively make up the dual rate, eq.~(\ref{dualrate}).  

For still higher masses (above $M\ge 1.5$~GeV) the most important 
hadronic contribution  arises from the associated semileptonic decay 
of charmed mesons ($D , \bar D \to e^+\nu X , e^-\bar\nu X$) which 
originate from primordial production of $c\bar c$ pairs.  A comprehensive  
evaluation of its features has been undertaken in ref.~\cite{GMRV96},  
as well as in the follow-up studies in \cite{Shu97,LVW98}.   
We will therefore not attempt to recalculate it but include the results
and arguments given in these references in the discussion of the 
spectra below.  

Finally, at masses starting from 1~GeV (primordial) Drell-Yan (DY) 
annihilation constitutes an important source of dileptons. Again, 
a detailed computation  
of DY-pairs at RHIC energies has been performed in ref.~\cite{GMRV96}. 
We here restrict ourselves to the case of central collisions 
of two equal nuclei with mass number $A$,   
\beq
\frac{dN_{DY}^{AA}}{dM dy}(b=0)=\frac{3}{4\pi R_0^2} \ A^{4/3} \
\frac{d\sigma_{DY}^{NN}}{dM dy} \ , 
\label{dyaa}
\eeq
in terms of the (standard) elementary Drell-Yan cross section in a 
nucleon-nucleon collision, 
\beq
\frac{d\sigma_{DY}^{NN}}{dM dy}=K \ \frac{8\pi\alpha^2}{9sM} 
\sum\limits_{q=u,d,s} e_q^2 \ \left[ q(x_1) \bar q(x_2) + 
\bar q(x_1) q(x_2) \right]  \ . 
\label{dynn}
\eeq
The (anti-) quark distribution functions $q(x_{1,2})$ ($\bar q(x_{1,2})$)  
are taken from the leading order (LO) GRV-94 
parameterization~\cite{GRV95} (including explicit isospin asymmetries)
in connection with a $K$-factor of 1.5 extracted from $p$-$A$ 
data~\cite{Spiel98}. Their arguments are related to the
center-of-mass rapidity $y$ and invariant mass of the lepton
pair as  $x_{1,2}=x e^{\pm y}$ with $x=M/\sqrt{s}$, where $s$ denotes
the total $cms$ energy of the nucleon-nucleon collision.
In eq.~(\ref{dyaa}), the root-mean-square radius
parameter $R_0\simeq1.05$~fm arises from a folding over a Gaussian
thickness function~\cite{Wong84}.
For (slightly) off-central collisions as considered below we simply
replace $A$ in eq.~(\ref{dyaa}) by the number of participants. 
Effects of nuclear shadowing have been estimated to reduce the total 
DY-yield in the RHIC regime by 20-50\%~\cite{GMRV96}, which is
within a similar uncertainty as implicit in the thermal production rate, 
eq.~(\ref{dualrate}).  

\section{Dilepton Production for $M$=0.5-3~GeV at RHIC}
\label{sec_spec}

\subsection{Thermal Fireball Model}
The total dilepton yield from thermal radiation in a heavy-ion collision 
obviously depends on the space-time history of the system, which
the thermal rates have to be integrated over. 
Here we employ a simplified thermal fireball description including, 
however, the most important features of a full hydrodynamic simulation
in a transparent way. Assuming a spatially averaged (isotropic) 
temperature-density profile at each time incident, 
the time- and momentum-integrated thermal mass spectrum
can be written as
\beq
\frac{dN_{ee}^{th}}{dM}=
\int\limits_0^{t_{fo}} dt \ V_{FC}(t) \int \frac{M d^3q}{q_0} \
\frac{dR_{ee}}{d^4q}[M,\vec{q};T(t),\mu_\pi(t),\mu_B(t)] \ 
{\rm Acc}(M,\vec q; y_0) \ .  
\eeq
Experimental acceptance corrections are encoded in the factor 
Acc$(M,\vec q; y_0)$, which is prescribed by the 
specific detector characteristics, and depends on the kinematics
of the produced pair as well as the rapidity positioning of the
thermal fireball ($y_0=0$ at RHIC). The latter has to be chosen in 
accord with measured hadronic rapidity distributions. As thermal 
fireballs typically imply smaller spreads in particle rapidity 
than found in heavy-ion reactions at ultrarelativistic 
energies~\cite{Stach96},  
multiple fireballs have to be used. At the full SpS energy, \eg, 
the full-width-half-maximum of negatives, $dN_{h^-}/dy$, amounts to about
$\Gamma_y^{\rm SpS}\simeq 4$. For thermal pion distributions at the
relevant temperatures one has $\Gamma_y^{th}\simeq 1.5-2$, so that 
two fireballs
are minimally required for SpS conditions, see, \eg, ref.~\cite{RW99}.  
At RHIC energies, a larger number would be necessary for a complete
description. However, since our focus here is on a comparably
small window around midrapidity ($\Delta y=\pm 0.35$ as covered by the 
two electron arms of the PHENIX detector), a single fireball suffices.  
The time-dependent volume is modeled by a cylindrical expansion as  
\beq
V_{FC}(t)= (z_0+v_z t) \
\pi \ (r_0 + \frac{1}{2} a_\perp t^2)^2 \ , 
\label{volexp}
\eeq
where $r_0$ is the initial transverse overlap of the two colliding
nuclei (determined by the impact parameter $b$), $v_z=c$ the 
longitudinal expansion velocity (reflecting the approximate rapidity 
coverage), and $a_\perp=0.03~$fm$^2/c^3$ is chosen to give a final 
transverse flow velocity of $v_\perp\simeq0.6c$ in connection with a 
typical freezeout time of $t_{fo}\simeq 20$~fm/c. The largest 
sensitivity of our results is attached to the initial longitudinal
size $z_0$ being equivalent to a formation time (or initial temperature),
which will be explicitly addressed below. 
 
\subsubsection{Hadronic Phase}
\begin{figure}
\vspace{-1cm}
\bce
\epsfig{file=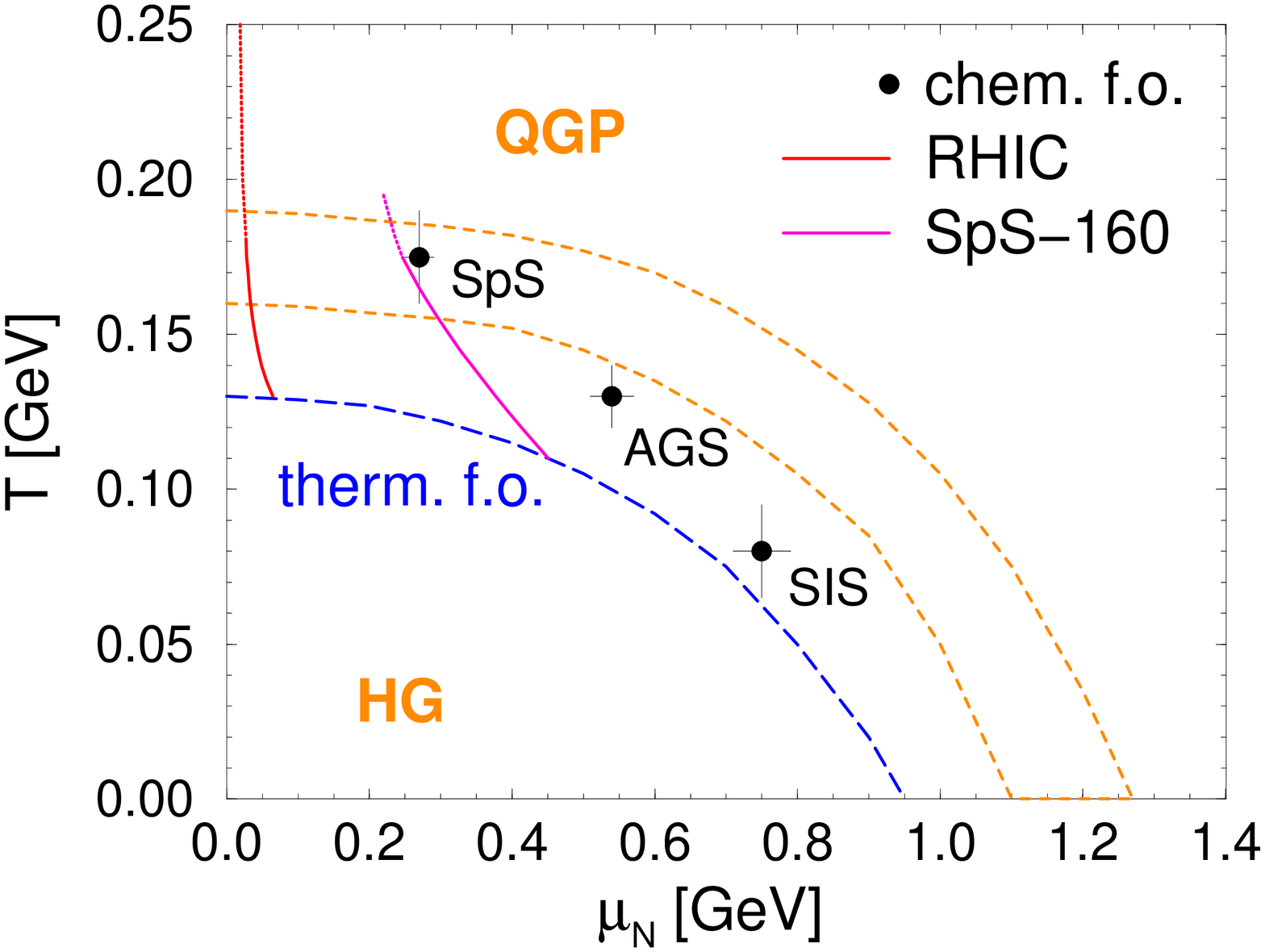,width=7.2cm,angle=-90}
\epsfig{file=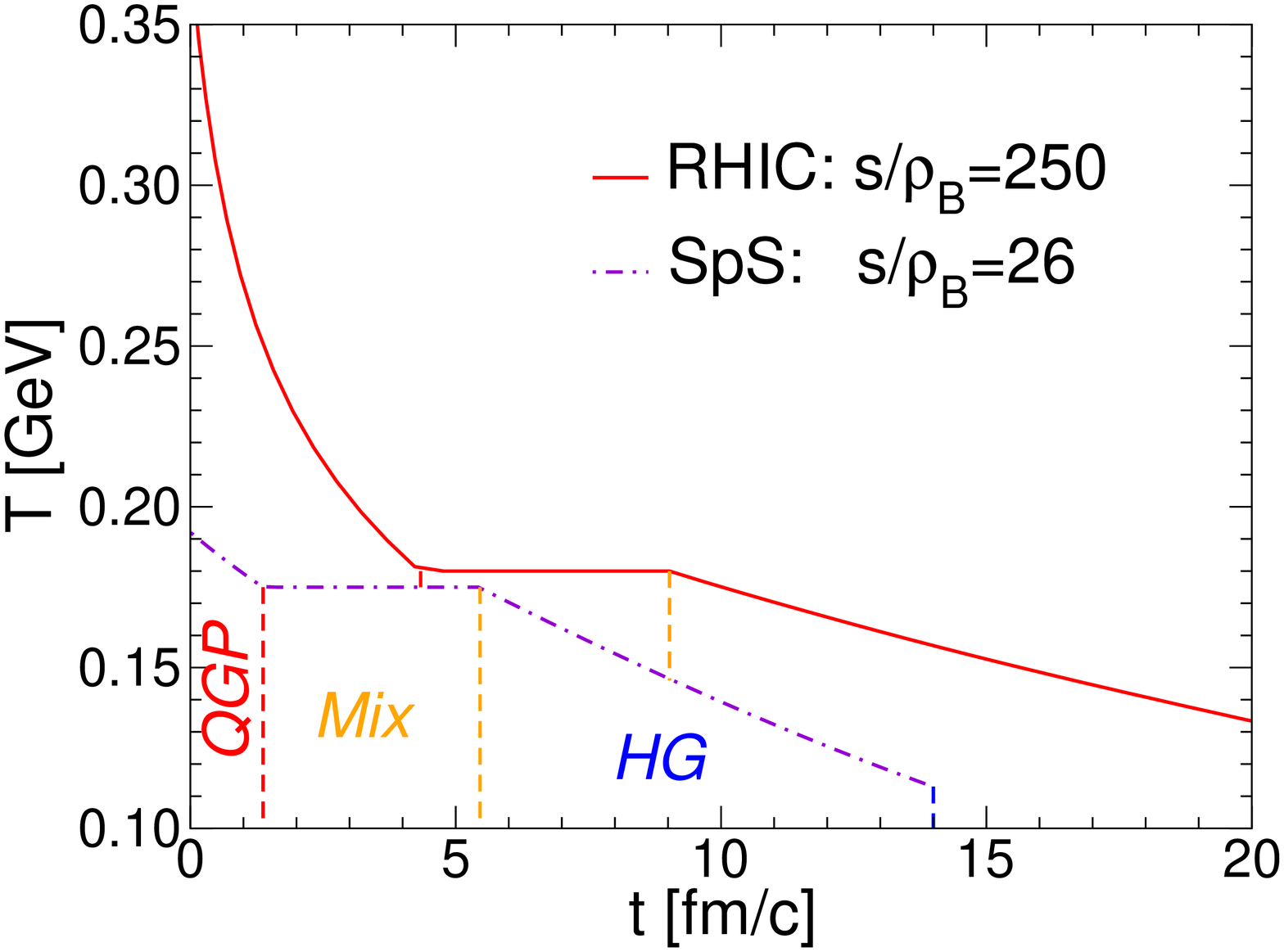,width=7cm,angle=-90}
\ece
\caption{Thermal fireball model description of central heavy-ion
collision at SPS and RHIC energies; 
left panel: Thermodynamic trajectories for isentropic
expansion in the temperature and nucleon-chemical-potential plane 
with additional $\pi$- and $K$-number conservation in the hadronic phases;
the short-dashed lines enclose a conjectured region for the phase boundary
between hadronic and quark-gluon matter, the long-dashed line represents
an estimate for the thermal (kinetic) decoupling conditions in heavy-ion
reactions, and the dots are the chemical freezeout points as extracted
in a thermal model analysis for fixed target experiments at the SpS 
(160~AGeV)~\protect\cite{pbm}, AGS (11~AGeV)~\protect\cite{pbm}
and SIS (2~AGeV)~\protect\cite{CR99};   
right panel: time-projected temperature evolution from the formation time
to thermal freezeout for central Au+Au (Pb+Pb) collisions at 
RHIC (SpS).} 
\label{fig_muTevo}
\end{figure}

To determine the thermodynamics of the fireball we employ the
picture of subsequent chemical and thermal freezeout, which has 
proven very successful in describing many observables  in heavy-ion
collisions over a wide range of energies (from SIS to SpS), such as
particle multiplicities~\cite{pbm,CR99}, flow, HBT radii, etc.. 
From the chemical freezeout point on 
the composition of hadrons which are stable on the
time scale of the fireball lifetime does no longer change. In systems
with relatively low baryon content (SpS and RHIC) it practically    
coincides with the critical temperature
of the chiral/deconfinement transition. Thus, for RHIC we fix it 
at $T_{ch}\equiv T_c\simeq 180$~MeV. The two necessary input numbers are
then the rapidity densities of charged particles and {\em net} baryons; 
dynamical models suggest, \eg,  $dN_{ch}/dy\simeq 800$ and 
$dN_{B}^{net}/dy\simeq 20$ at midrapidity for the full RHIC energy of 
$\sqrt{s}=200$~AGeV~\cite{qm99} (the charged particle multiplicity is
in line with a naive extrapolation based on first RHIC measurements at 
lower energies~\cite{phobos00}). With a standard hadronic resonance 
gas equation of state (EoS) these numbers imply an entropy per baryon 
of $s/\varrho_B^{net}=250$, corresponding
to a baryon chemical potential of $\mu_{B}^{ch}=27$~MeV at $T_{ch}$.   
Isentropic expansion (\ie, at constant $s/\varrho_B^{net}$) then governs 
the evolution of $(T,\mu_B)$ towards thermal (kinetic) freezeout, 
cf.~left panel of fig.~\ref{fig_muTevo}.   
Since pion and kaon numbers are, by definition, also conserved,
pertinent (effective) chemical potentials $\mu_\pi$ and $\mu_K$ are
simultaneously incorporated. 
As opposed to SpS conditions, their values at thermal freezeout for 
RHIC are quite small reaching $\mu_{\pi}^{th}\simeq 10$~MeV, 
$\mu_{K}^{th}\simeq 35$~MeV at $T_{th}\simeq 130$~MeV\footnote{The 
reason for that is the larger 
number of baryonic resonances at SpS, which serve as a storage of
pion-number in the earlier stages, being released into  pionic degrees
of freedom towards thermal freezeout.}.  
Nevertheless, we include their enhancement effect in the thermal  
dilepton production rate from the hadronic phase as well as in the
cocktail contribution, eq.~(\ref{cktl}), via 
\noindent
additional fugacity factors, which
are $z_\pi^2$ for $\rho$ decays, $z_\pi^3$ for $\omega$ decays,
$z_K^2$ for $\phi$ decays, and $z_\pi^4$ in the IMR
corresponding to $\pi a_1$ and $\pi\omega$ annihilation
($z_x=\exp[\mu_x/T]$).

We finally quote the total and net baryon densities at chemical freezeout 
for the fireball evolutions shown in Fig.~\ref{fig_muTevo}: at RHIC one 
has $\varrho_{B,ch}^{net}=0.14\varrho_0$ 
and $\varrho_{B,ch}^{tot}=0.54\varrho_0$,
whereas at SpS $\varrho_{B,ch}^{tot}=1.5\varrho_0$ and
$\varrho_{B,ch}^{net}=1.41\varrho_0$.

\subsubsection{Quark-Gluon Plasma and Mixed Phase}
For the early stages (starting at the formation time $\tau_0$) 
we assume an ideal QGP EoS characterized by an entropy density
\beq
s_{QGP}= d_{QG} \ \frac{4\pi^2}{90} \ T^3 \  
\label{sqgp}
\eeq
(with a quark-gluon degeneracy $d_{QG}=16+10.5 N_f$ and $N_f=2.5$), 
ignoring effects from a net baryon number ($\mu_q\ll T$). 
Since most of the entropy is presumably produced in the 
pre-equilibrium stages of a collision with little changes
once thermal equilibrium is established~\cite{ER00}, we can extrapolate
from the chemical freezeout backward into the QGP phase using isentropes
as before.  At $T_c$, the fraction $f$ of matter
in the hadronic phase can be inferred from the standard entropy balance
\beq
f \ s_{HG}(T_c) + (1-f) \ s_{QGP}(T_c) = S_{tot}/V_{FC}(t) \  ,   
\label{balance}
\eeq 
while for $s(t)=S_{tot}/V_{FC}(t)>s_{QGP}(T_c)$ the entire
system is in the plasma phase. 

\begin{figure}
\vspace{-0.7cm}
\bce
\epsfig{file=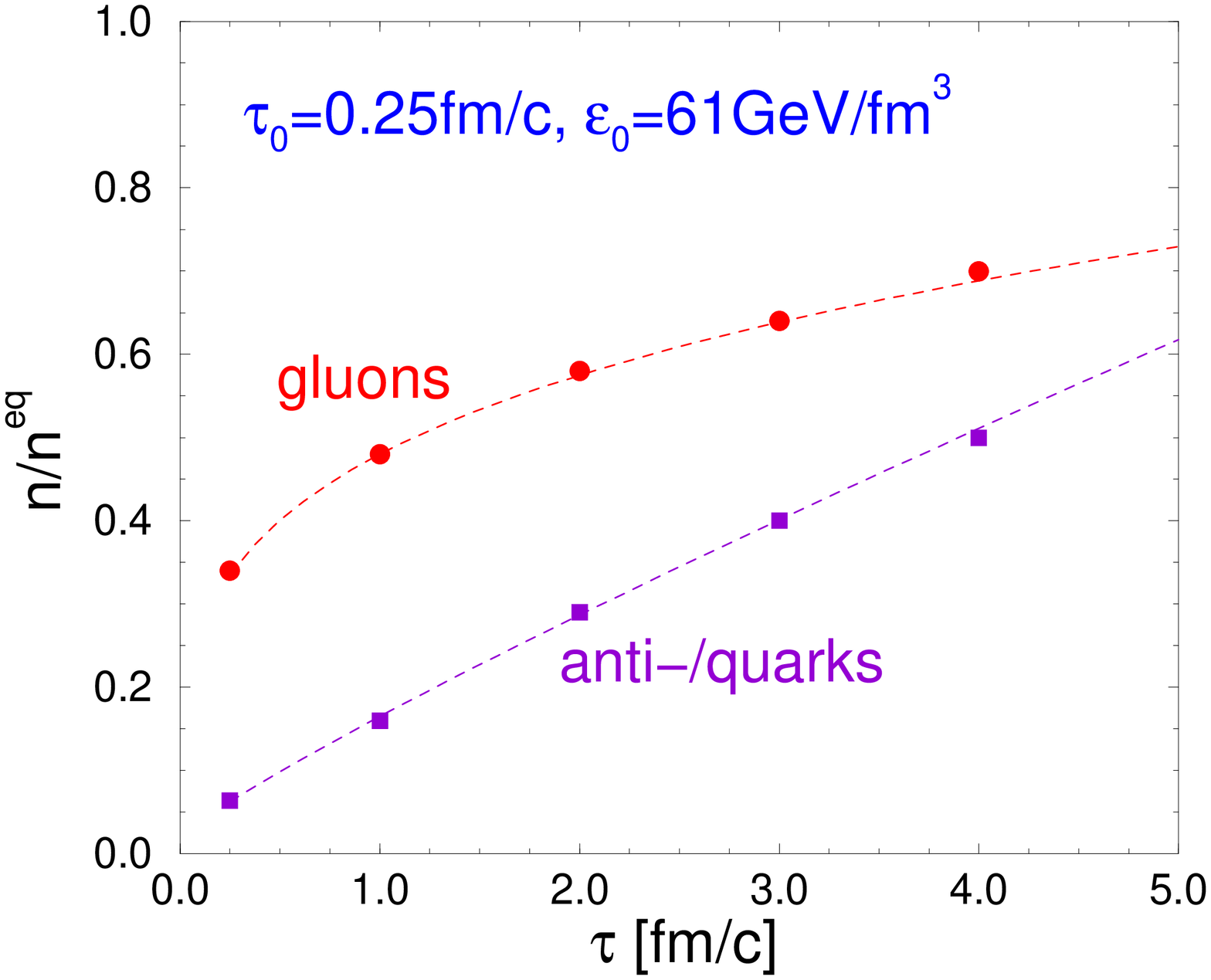,width=7.5cm,angle=-90}
\ece
\caption{Time evolution of anti-/quark (squares) and gluon (dots)
fugacities as computed in a hydrodynamic simulation~\protect\cite{ER00}
based on initial conditions obtained from a parton cascade 
model~\protect\cite{sspc}. The dashed lines show the results of 
the analytic fit, eqs.~(\protect\ref{fugglu}) and (\protect\ref{fugqrk}).}
\label{fig_fugqgp}
\end{figure}

At collider energies (RHIC and LHC) gluons are expected to play the 
dominant role in early entropy production and 
equilibration processes~\cite{Shu92}, due to both their larger 
color charge as well as the steep increase of their structure functions 
towards low (Bjorken-) $x$. Although thermal equilibration seems to be
achieved at a formation time of well below 1~fm/c, the chemical
abundances of gluons and especially (anti-) quarks are likely to be
strongly {\em under}-saturated. {\it E.g.}, in the 'self-screened 
parton cascade' model (SSPC) the initial parton fugacities at the thermal
formation time $\tau_0=0.25$~fm/c amount to 
$\lambda_g(\tau_0)=n_g(T_0)/n_g^{eq}(T_0)=0.34$ and  
$\lambda_{q,\bar{q}}(\tau_0)=n_{q,\bar{q}}(T_0)/n_{q,\bar{q}}^{eq}(T_0)
=0.064$~\cite{sspc} 
(estimates based on perturbative QCD are even lower, see, \eg, \cite{EsKa}).  
To study the impact of such a scenario on the dilepton radiation from
the plasma, we employ the results of ref.~\cite{ER00}, where the SSPC
initial conditions have been evolved using hydrodynamic rate equations until 
the onset of hadronization. For our purposes we parameterize the time
dependence of the fugacities as found there by
\bea
\lambda_g(\tau)&=& 0.48~\tau^{0.26}
\label{fugglu}
\\
\lambda_{q,\bar{q}}(\tau)&=& 0.145~\tau^{0.88} + 0.02 
\label{fugqrk}
\eea
($\tau$ in [fm/c]), cf. fig.~\ref{fig_fugqgp}. 
Entropy production after $\tau_0$ was found to be
on the 10\% level which will be neglected here. Thus we can calculate
the temperature (needed for the dilepton production rate) from
the entropy density via
\bea
s_{QGP}^{off}(t) &=& \left[ 16\lambda_g(t)+10.5 N_f \lambda_q(t)\right] \ 
\frac{4\pi^2}{90} \ T^3 
\nonumber\\
&\equiv& d_{QG}(t) \ \frac{4\pi^2}{90} \ T^3 \ .
\label{sqgpoff}
\eea
Since chemical equilibration is not attained before entering  
the mixed phase, the  construction of the latter has to be modified.
We assume its onset to be determined by the same critical entropy
density as in the equilibrium scenario so that the critical
temperature 
\beq
T_c^{QGP}(t)=\left[\frac{90~s_{QGP}^c}{4~d_{QG}(t)~\pi^2}\right]^{1/3}
\eeq
acquires a time dependence while for the hadron gas $T_c^{HG}=180$~MeV
is kept fixed.  Eq.~(\ref{balance}) still serves to obtain the volume
fraction of the hadron gas, $f(t)$.      

As a consequence of the quark-undersaturation, the equilibrium dilepton 
production rate from $q\bar{q}$ annihilation, eq.~(\ref{dualrate}),
is (approximately) proportional to a fugacity factor 
$\lambda_q \lambda_{\bar q}$, which translates into a strong suppression 
of the thermal yield. However, if the same initial entropy density is
created as in the equilibrium scenario (which largely governs the
finally observed number of secondaries), this suppression 
will be (partially) compensates by higher temperatures in the 
thermal Bose factor of the virtual photon~\cite{SX93}. Also, if the 
gluon fugacities are substantially larger than the anti-/quark ones, 
$\alpha_s$-corrections to the $q \bar{q}$ rate become
relatively more important, \eg, through the process $qg\to qee$. As 
shown in ref.~\cite{LK00} related contributions are comparable to 
$q\bar q$ production in a more extreme scenario of undersaturation with 
$\lambda_g/\lambda_q\simeq 10$. Since in our case 
$\lambda_g/\lambda_q < 3$ after $\tau=1$~fm/c (and quickly 
decreasing further), we keep ignoring $\alpha_s$-corrections (as done
for the equilibrium case) being aware of
possibly underestimating the thermal yield from the off-equilibrium 
plasma by a few tens of percent.

\subsection{Dilepton Spectra}
Let us now turn to the results for the space-time integrated 
dilepton spectra in central Au+Au collision at $\sqrt{s}=200$~AGeV.
Our standard parameter set for the volume expansion within the 
thermal fireball model, eq.~(\ref{volexp}), is chosen as
$v_z=c$, $r_0=6.5$~fm (reflecting an impact parameter $b\simeq 1$~fm), 
$a_\perp=0.03$~fm$^2$/c$^3$ as well as initial longitudinal 
size $z_0=0.6$~fm (translating
into an initial temperature of $T_0=370$~MeV for a chemically 
equilibrated QGP) and total fireball lifetime of $t_{fo}=20$~fm/c
(resulting in a thermal freezeout temperature of 
$T_{th}=133$~MeV\footnote{note that this value is about 10-15\%
larger than at SpS energies owing to the substantially smaller
baryon content at RHIC}). 
 
\begin{figure}
\vspace{-0.3cm}
\bce
\epsfig{file=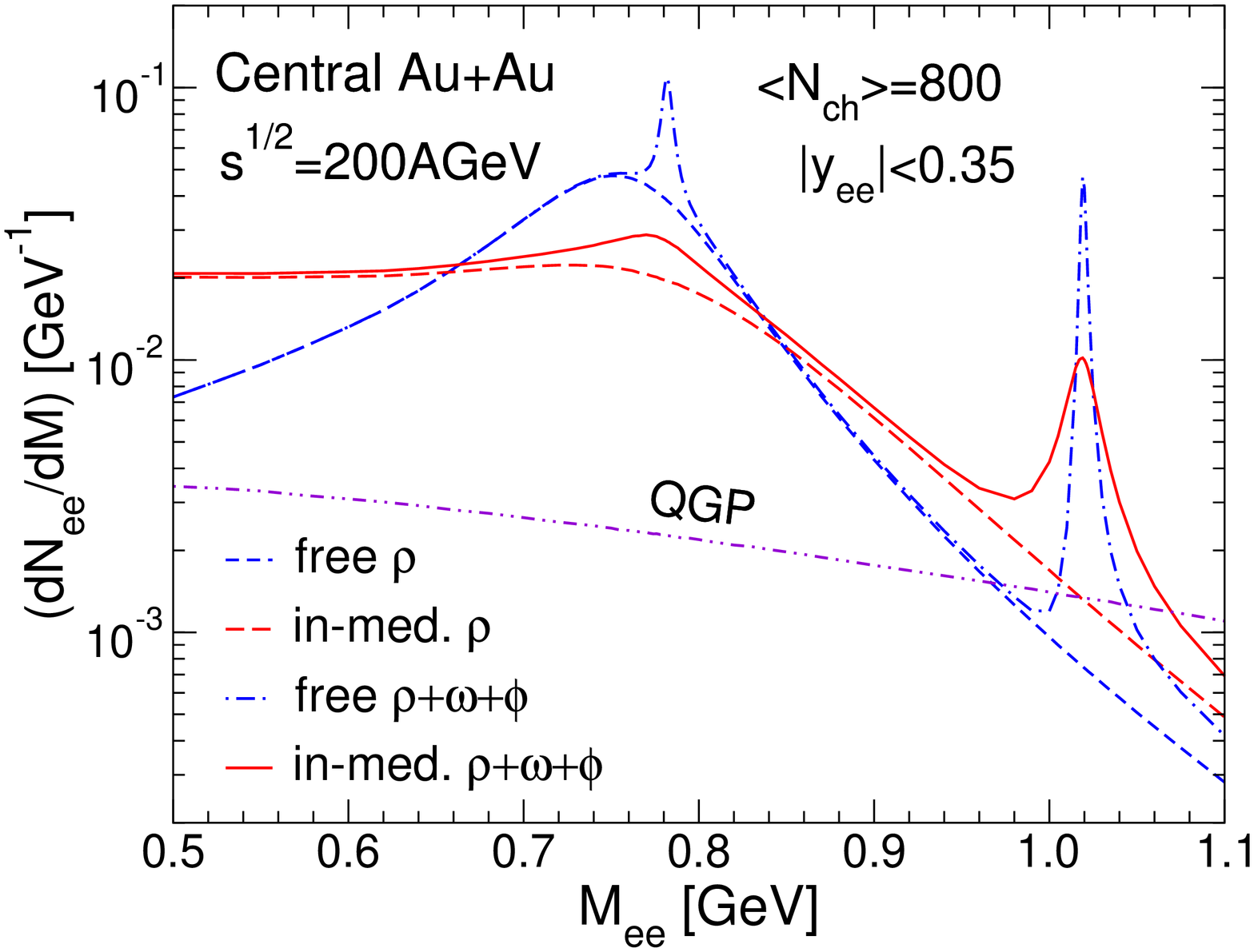,width=7.3cm,angle=-90}
\hspace{-1.4cm}
\epsfig{file=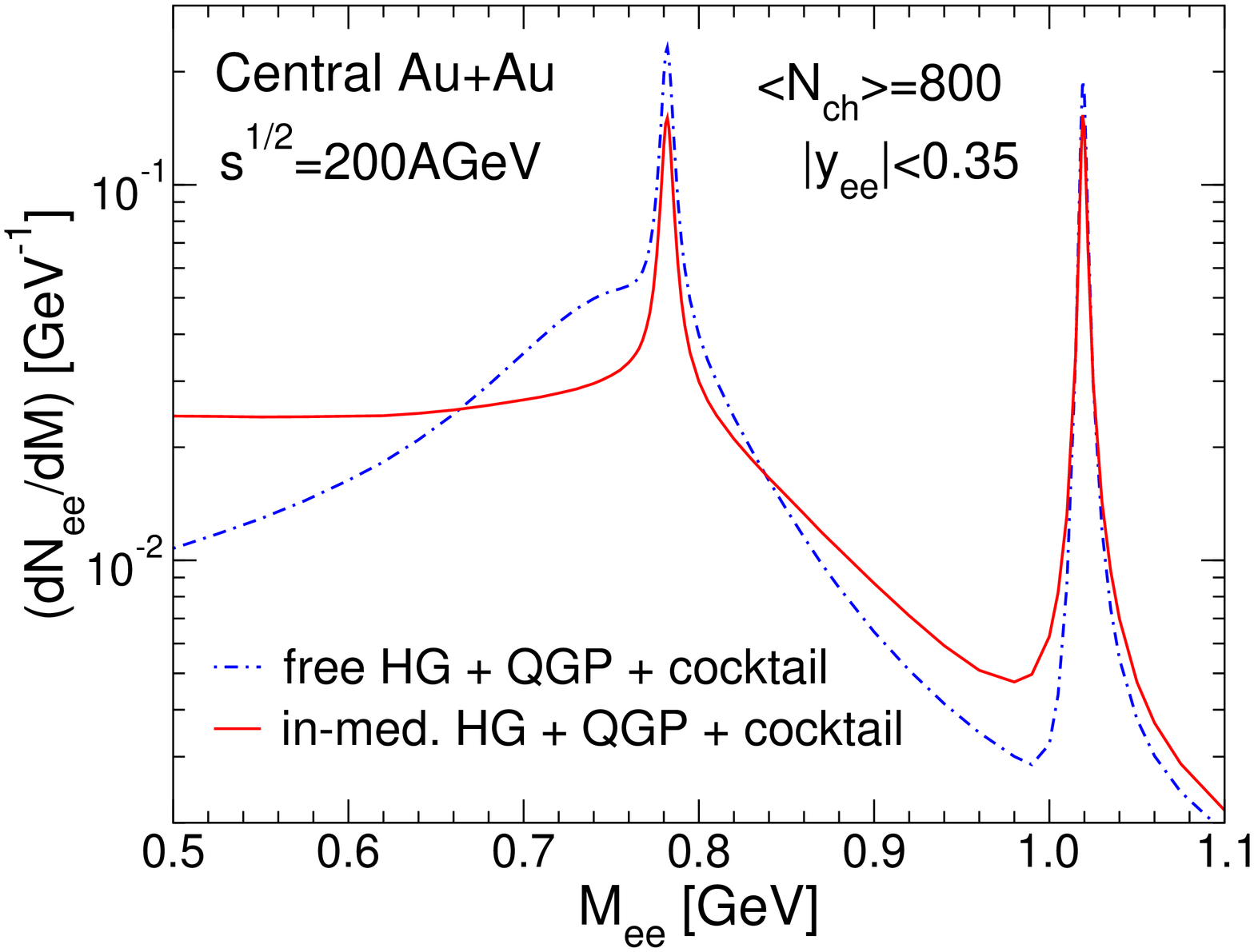,width=7.3cm,angle=-90}
\ece
\caption{Low-mass dilepton spectra around midrapidity in $b=1$~fm Au+Au 
collisions at RHIC energies.  
Left panel: thermal radiation from vector-meson decays in the hadronic
phase with and without medium effects in the pertinent spectral functions, 
as well as from perturbative $q\bar q$ annihilation in the plasma
phase (dashed-double-dotted line). 
Right panel: combined yields from the hadronic and plasma phase as well
as from $\omega, \phi\to ee$ decays after freezeout; full and dashed-dotted
lines correspond to the calculation with and without medium modifications, 
respectively,   in the $\rho$, $\omega$ and $\phi$ spectral functions.}
\label{fig_speclmr}
\end{figure}

We first focus on the LMR, cf.~fig.~\ref{fig_speclmr}. Only lepton pairs 
which fall into the rapidity range $-0.35<y<0.35$ are included, 
integrated over all transverse momenta. 
The main significance of in-medium effects in the radiation from 
the thermal fireball (left panel of fig.~\ref{fig_speclmr}) 
are a (strong) suppression of the yield in the $\rho$-$omega$
region and an enhancement below M=0.65~GeV  as well as above
0.85~GeV. The $\phi$, on the other hand, survives as a pronounced, 
albeit broadened,  resonance structure.
These features become less distinct once the 'cocktail' contribution
(from $\omega$ and $\phi$ decays) and the smooth yield
from the QGP (which constitutes around $\sim$20\%) are added. Between
$M=0.7$~GeV and 0.8~GeV
one still observes an appreciable $\sim$50\% reduction
caused by in-medium effects (and a factor $\ge$~2 enhancement towards
$M=0.5$~GeV). Experimentally it might be even possible to subtract
the 8~MeV wide contribution of free $\omega$ decays and then try
to extract a remaining structure of $\sim$50~MeV width -- originating
from $\omega$ decays within the interacting fireball -- which is still
well below the free $\rho$ width.
In the $\phi$ region our results suggest that such a procedure is
less promising to succeed. The factor of 2 increase just below
1~GeV might be more easy to identify.

\begin{figure}
\vspace{-0.3cm}
\bce
\epsfig{file=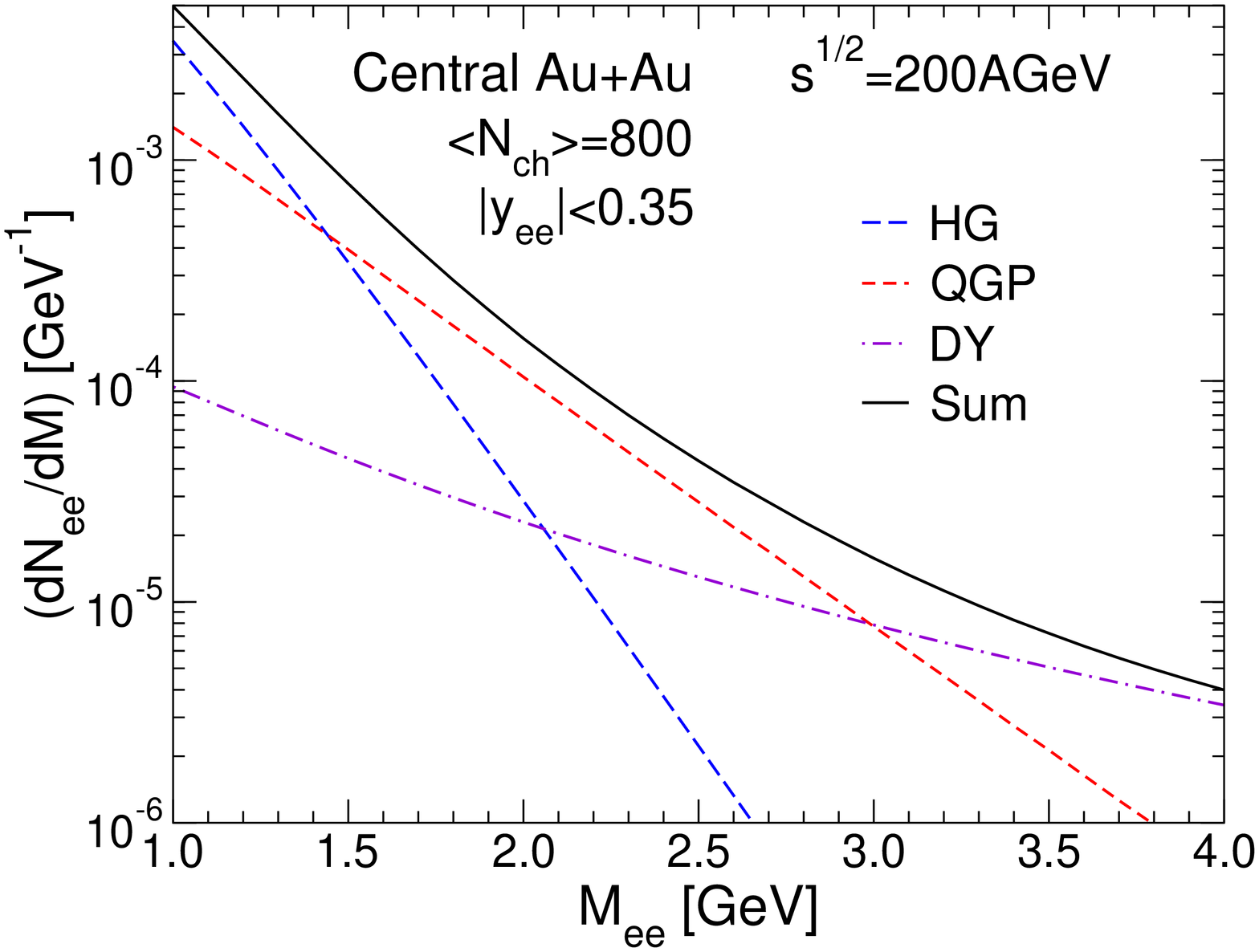,width=7.3cm,angle=-90}
\hspace{-1.4cm}
\epsfig{file=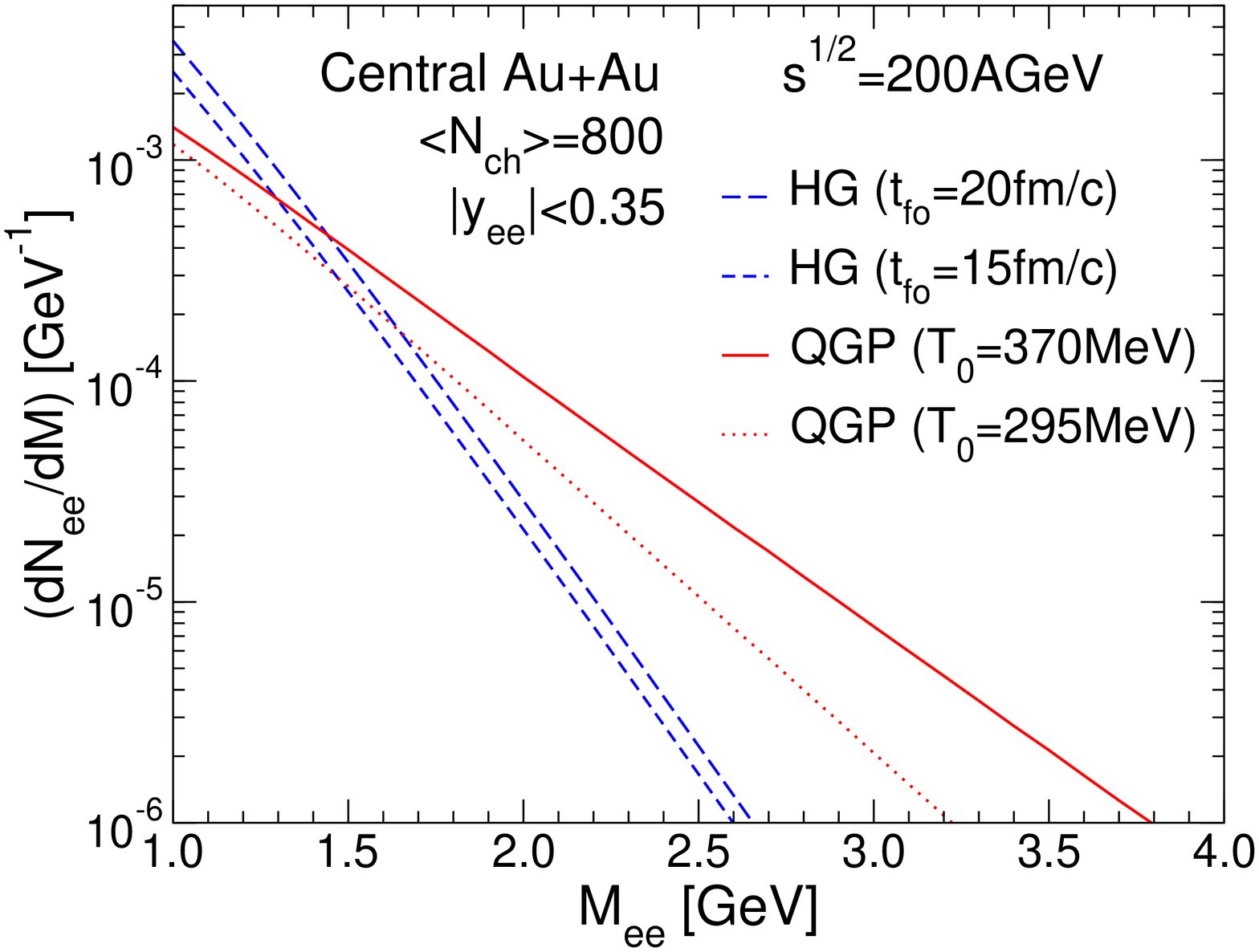,width=7.3cm,angle=-90}
\ece
\caption{Intermediate-mass dilepton spectra at RHIC energies around
midrapidity assuming a chemically equilibrated QGP throughout its
lifetime.
Left panel: Decomposition of the thermal fireball radiation
(with $T_0=370$~MeV, $t_{fo}=20$~fm/c) into QGP (short-dashed
line) and hadron gas (long-dashed lines) parts (including their respective
yields from the mixed phase), compared to Drell-Yan annihilation (dashed-dotted
line) and the total sum (note that open charm decays are not accounted for).
Right panel: Influence of the initial temperature  and fireball lifetime
on the plasma and hadron gas radiation, respectively.}
\label{fig_specimr}
\end{figure}

Next, we move to the IMR, first assuming QGP formation and evolution
in full chemical equilibrium, cf.~fig.~\ref{fig_specimr}. From the left
panel one sees that the hadron gas radiation prevails until
$M\simeq 1.5$~GeV
before the thermal radiation from the QGP phase takes over due to
its smaller slope parameter (originating from larger temperatures).
The QGP signal  continues to outshine the (primordial) Drell-Yan yield
up to about $M\simeq 3$~GeV (similar features have been found
in refs.~\cite{KKLM86,CFR87}). Taken together, the total sum exhibits a
noticeable change in slope between 1 and 4~GeV.

\newpage

Not included in fig.~\ref{fig_specimr} are the contributions from
semileptonic decays of correlated $c\bar c$ pairs.
Extrapolations of their production cross sections in $pp$
collisions to Au+Au collisions at RHIC lead to rather large dilepton
emission, exceeding thermal radiation estimates by up to an order of 
magnitude~\cite{GMRV96,GKP98} at invariant masses $M\simeq 2$~GeV.
However, subsequent analyses~\cite{Shu97,LVW98,GKP98} investigated
energy-loss effects for the charm quarks when propagating through hot 
and dense matter~\cite{GW94,BDPS95}.
It was shown that the (rather hard) primordial transverse momentum
distributions are substantially softened,  becoming
essentially thermalized. Consequently, the pertinent lepton pair
invariant mass spectrum is strongly depleted above $M\simeq 2$~GeV, by
about two orders of magnitude, thus dropping below the Drell-Yan
yield\footnote{Most of the open charm contribution is redistributed to
$M\le 1$~GeV; this reflects the kinematics of the two leptons
emerging in randomly oriented semileptonic 3- and 4-body
decays of $c(\bar{c})\to l^\pm \bar{\nu} (\nu) X$,
where the (thermal) charm quarks carry little
kinetic energy so that $E_l\simeq m_c/3\simeq 0.5$~GeV.}. In addition,
one might be able to assess the correlated charm decays independently
through $e\mu$ coincidence measurements~\cite{phenix93} or
secondary vertex displacements~\cite{priv00}.

To test the sensitivity of our results with  
respect to the fireball evolution we have displayed in the right
panel of fig.~\ref{fig_specimr} the outcome of the following
two calculations:   
first, doubling the initial longitudinal extent
(or formation time) to $z_0=1.2$~fm/c (which translates into a reduction
of the initial temperature to $T_i=295$~MeV), the QGP radiation is
appreciably decreased (by $\sim$50\%). In turn, the hadron gas evolution 
and corresponding dilepton yield is practically
identical to the $z_0=0.6$~fm case. Second, doubling the transverse
acceleration to $a_\perp=0.06$ together with a reduced lifetime of 
$t_{fo}\simeq 15$~fm/c (to maintain a similar freezeout temperature 
of $T_{th}\simeq 135$~MeV as before) 
reduces the hadron gas part accordingly (by $\sim$25\%), whereas the 
plasma yield is essentially unchanged. The latter can be attributed to  
the predominant longitudinal expansion in the early stages 
leaving little impact of an increased transverse acceleration
on the first 5~fm/c or so. 
Thus, not surprisingly, the magnitude of the plasma yield is closely
linked to the formation time, whereas the hadron gas yield is a rather
direct measure of the lifetime the system spends in the hadronic phase. 

\begin{figure}
\bce
\epsfig{file=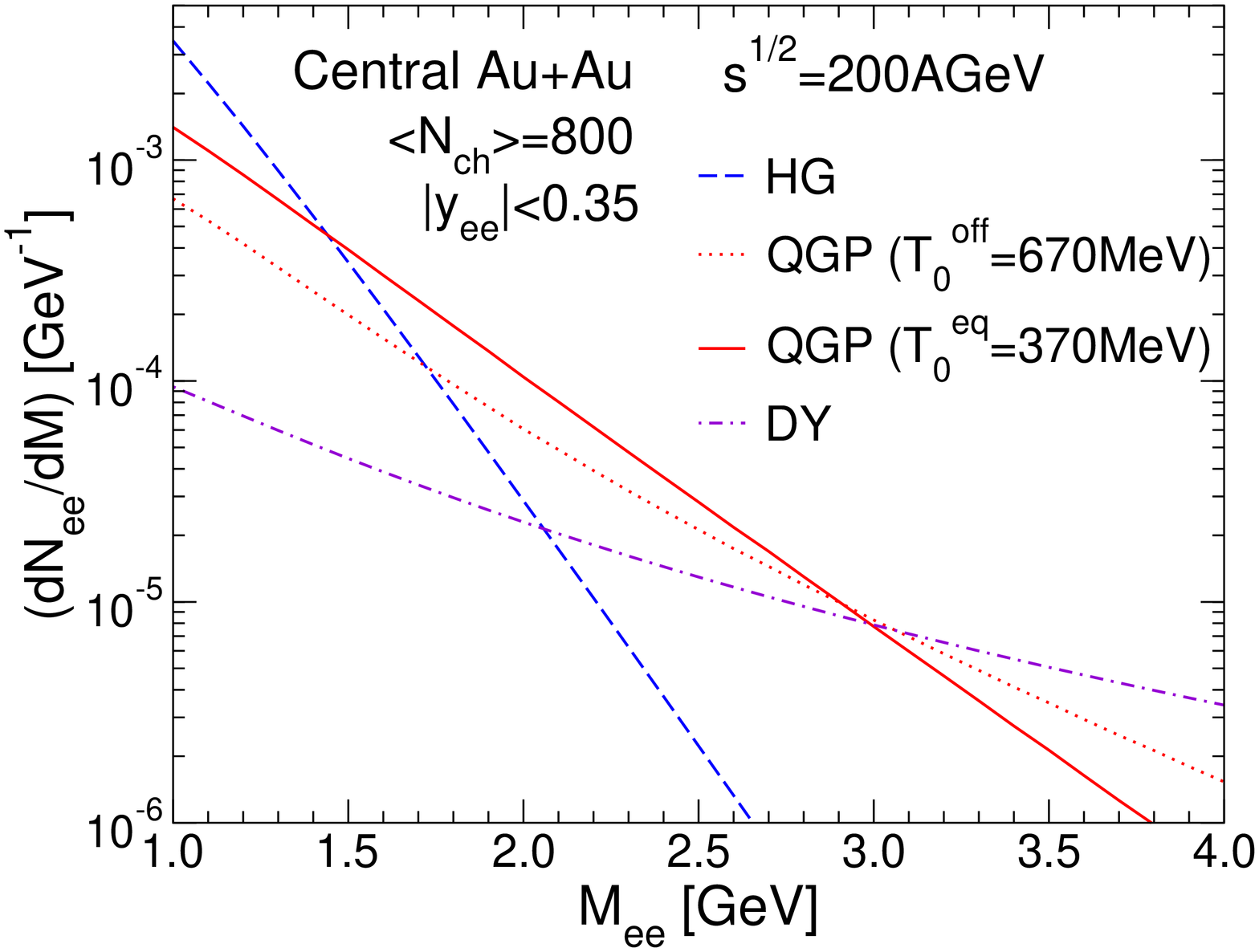,width=7.3cm,angle=-90}
\hspace{-1.4cm}
\epsfig{file=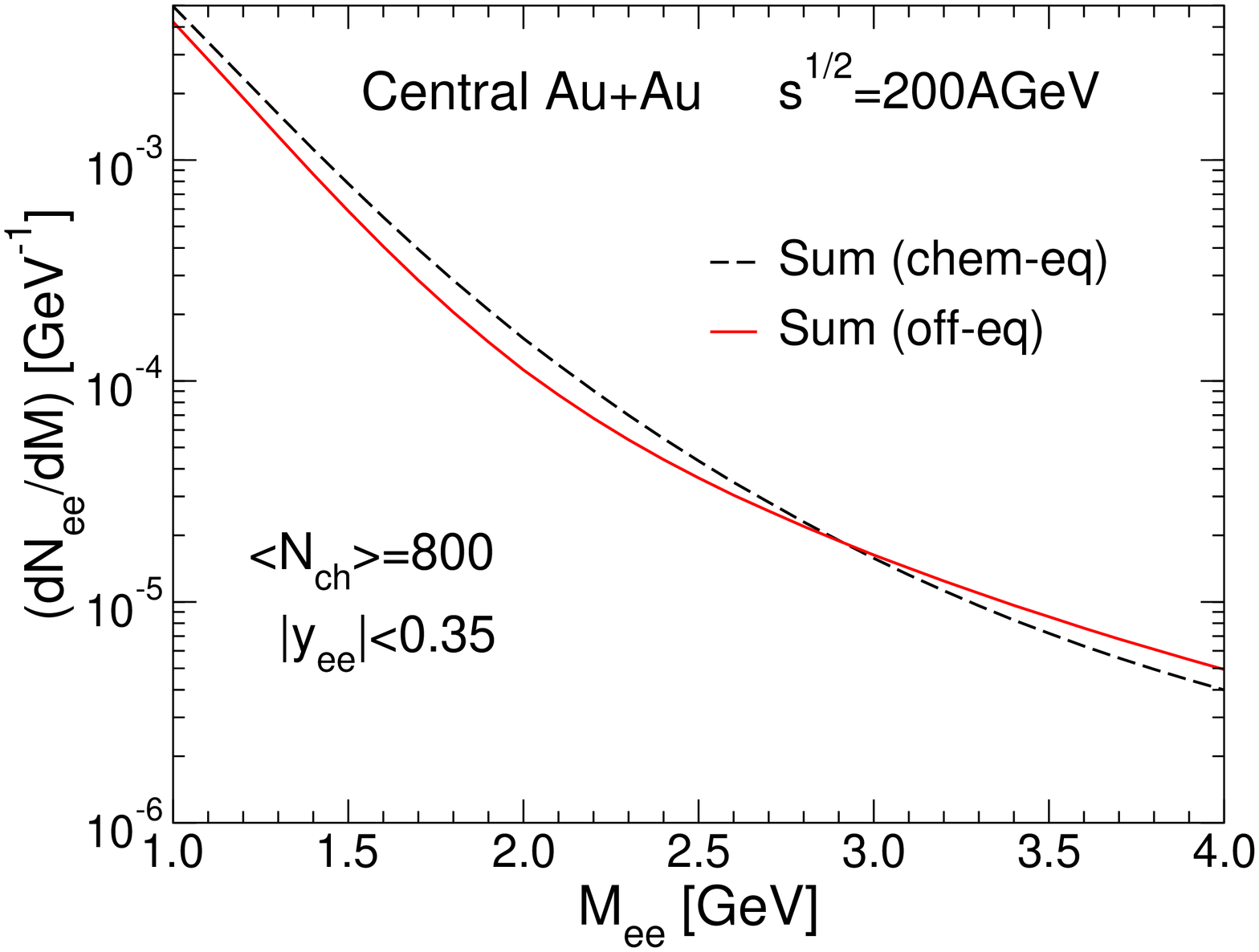,width=7.3cm,angle=-90}
\ece
\caption{Consequences of a chemically undersaturated QGP for IMR dilepton 
spectra at RHIC energies.  
Left panel: comparison of the plasma yield from a 
chemical equilibrium (short-dashed line) and off-equilibrium (full
lines) scenario adopted from refs.~\protect\cite{sspc,ER00}  
(the hadron gas contribution is virtually identical for both cases).  
Right panel: total sum of hadron gas, Drell-Yan and QGP contributions
with (full line) and without (dashed line) chemical undersaturation 
in the latter.}
\label{fig_specimroff}
\end{figure}

Furthermore, we address the question to what extent chemical 
undersaturation of thermal quark and gluon distribution functions in the 
early plasma stages affects the IMR dilepton spectra.  The time 
evolution of the fugacities (cf.~fig.~\ref{fig_fugqgp}) is implemented 
as described in the previous section (to account for the twice larger 
initial pressure we also used an accordingly increased transverse 
acceleration). Requiring equal initial entropy densities, the 
off-equilibrium scenario leads to a slightly reduced (enlarged) QGP yield 
below (above) $M\simeq 3$~GeV due to a harder slope caused
by the higher initial temperature\footnote{The small deviations from 
the results presented in ref.~\cite{Rapp00} arise from the fact that 
in there the entropy of the QGP was calculated with $N_f=3$ (rather 
than $N_f=2.5$ as here).}, see fig.~\ref{fig_specimroff}.    
This result is a little less optimistic than the one of 
Shuryak and Xiong~\cite{SX93}, who predict an enhancement
of thermal IMR dilepton radiation in a similar 'hot-glue' scenario 
(see also ref.~\cite{Raha90}).
On the other hand, Lin and Ko, using initial conditions estimated 
from perturbative-QCD~\cite{wang97}, find very strong 
suppression of the thermal yield which comes out well 
below Drell-Yan annihilation~\cite{LK00}. The employed initial conditions, 
$\lambda_{q,0}=\lambda_{\bar q,0}=0.0072$, $\lambda_{g,0}=0.06$ 
and $T_0=570$~MeV, correspond to an entropy density of 
$s_0\simeq 12$~fm$^{-3}$, which is an order of magnitude smaller 
than in the SSPC approach adopted here (eqs.(\ref{fugglu}), 
(\ref{fugqrk}), and (\ref{sqgpoff}) give $s_0^{off}=122$~fm$^{-3}$; 
note that for a fully equilibrated QGP at $T=T_c\simeq 180$~MeV one 
has $s=14$~fm$^{-3}$).

\begin{figure}
\vspace{-1cm}
\bce
\epsfig{file=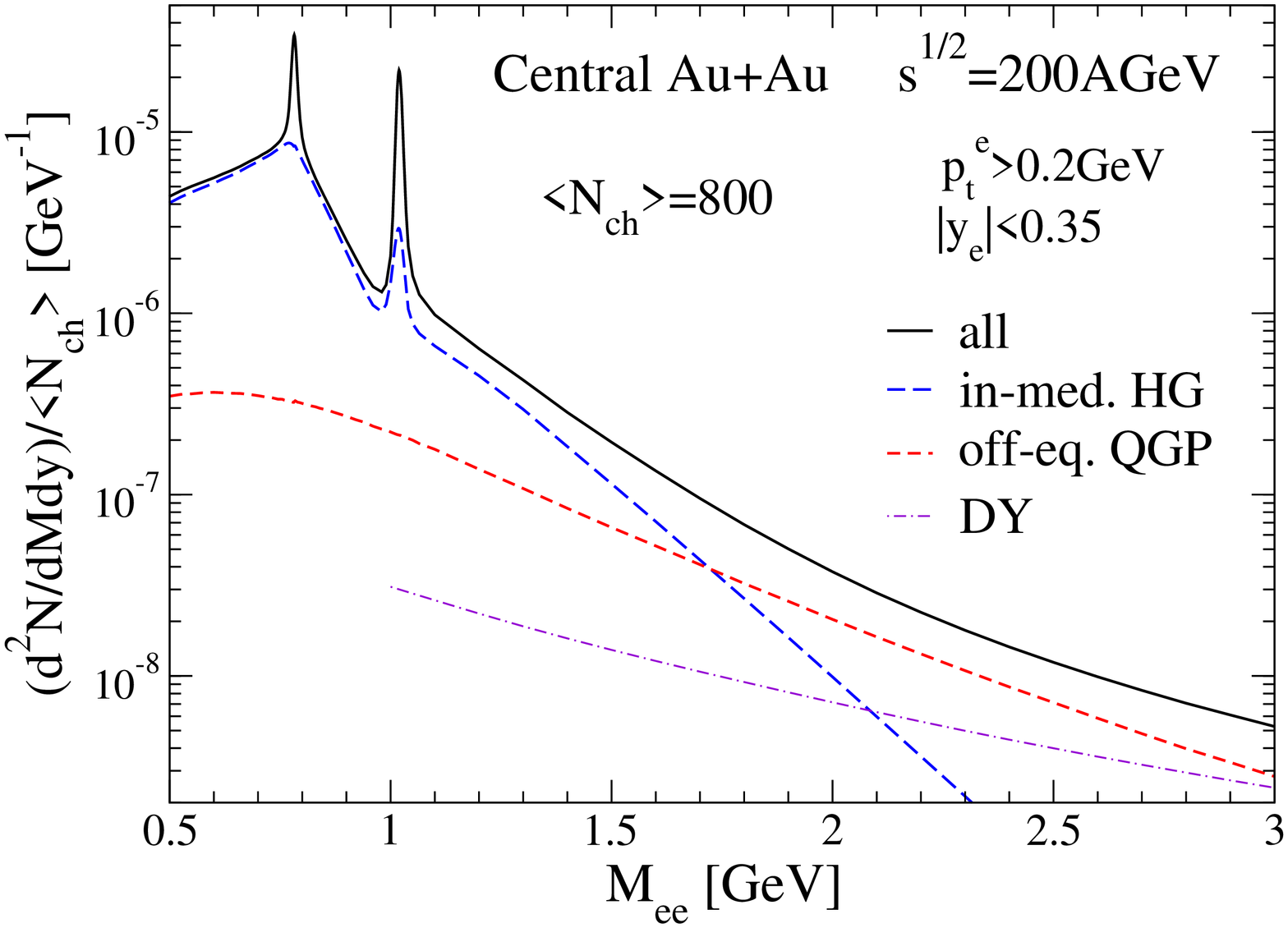,width=12cm,angle=-90}
\ece
\caption{Total dilepton spectrum from $b=1$~fm Au+Au collisions
at RHIC energies around midrapidity
including schematic experimental acceptance cuts appropriate for
the PHENIX experiment (the $\omega,\phi\to e^+e^-$ cocktail contributions
are not separately shown but included in the solid curve). 
Note that semileptonic decays of correlated anti-/charm and anti-/bottom 
quarks have been left out.}
\label{fig_spectot}
\end{figure}

Finally, in fig.~\ref{fig_spectot} we summarize the results for
the combined LMR and IMR
spectra with additional acceptance restrictions on the single-electron 
tracks as applicable for the PHENIX detector ($p_t^e>0.2$~GeV, 
$|y_e|\le 0.35$, opening angle $\Theta_{ee}\ge 35$~mrad and an 
overall pair mass resolution of $\sigma_M/M=0.5$~\%). 

\section{Summary and Conclusions}
\label{sec_concl}
	
Within a thermal framework we have discussed properties 
of dilepton radiation from ultrarelativistic heavy-ion collisions 
under conditions relevant for the PHENIX experiment at RHIC. 

In the first part of this article dilepton emission from static equilibrium 
matter has been evaluated. The emphasis was on the low-mass region
($M\le 1$~GeV), where it was found that medium modifications of the 
light vector mesons induce an appreciable reshaping of the free 
electromagnetic current correlator.  
Based on standard many-body techniques, hadronic interactions in hot
matter at small net-baryon content generate a strong broadening of 
both the $\rho$ and $\omega$ spectral functions. 
Close to $T_c$, the $\omega$ width increases by almost 
a factor of 20 over its vacuum value, with non-negligible contributions 
from rescattering off thermally excited anti-/baryons. The  
$\phi$ meson retains more of its resonance structure, although its free 
width is also augmented by up to a factor of 7-8.   
With that exception in mind, the resulting in-medium hadronic dilepton 
production rates approach the continuum shape of lowest-order 
$q\bar{q}$ annihilation to within a factor of 2 in the vicinity 
of the phase boundary. We have thus argued 
that towards $T_c$ 'quark-hadron duality' in the vector channel, 
which in vacuum is marked by the applicability of perturbative QCD 
above $M\simeq 1.5$~GeV, extents down to rather low masses of 
$\sim$0.5~GeV (this scale is in fact comparable to typical 
thermal energies of free (massless) partons). These findings comply with    
earlier analyses that were carried out for more baryon-rich matter. 

To calculate dilepton spectra for central Au+Au collisions at RHIC 
energies we have employed a thermal fireball model
including chemical off-equilibrium effects in both the QGP and
hadron gas phases. In the hadronic evolution from chemical to
thermal freezeout quite moderate  pion and kaon chemical potentials of 
10-30~MeV emerge.  The resulting low-mass dilepton yield reflects upon 
the lifetime of the hadronic system. As a promising signature, the final 
invariant-mass spectra contain a $\sim$50~MeV-wide structure from in-medium 
$\omega$ decays. This is well below the free $\rho$ width (150~MeV), but
much larger than the width associated with free $\omega$ decays (8.4~MeV). 
The latter, leading to copious dilepton production after freezeout, 
will mask the in-medium contribution in the measured spectra.  
However, with sufficiently high mass resolution as anticipated in the  
PHENIX experiment, the $\omega$ 'cocktail' yield should be subtractable, 
thus enabling access to the in-medium part.       
 
In the intermediate-mass region the focus was on radiation from a 
thermalized partonic medium. Here, our investigation was of a more 
qualitative nature.  
On the one hand, this is due to uncertainties in the QGP production 
rate (which we might have underestimated by up to a factor of 2)   
as well as in the formation/composition of the 'Gluon-Quark' plasma.  
Nevertheless, our main point in this context should not be affected, 
namely that chemical undersaturation 
scenarios for gluons and especially anti-/quarks in the early stages 
do not decrease the thermal radiation in the 1.5-3~GeV region 
below the Drell-Yan yield --  provided the production of the finally 
observed entropy (as encoded in produced secondaries)  
essentially  occurs within the first fraction of a fm/c after nuclear 
impact (parton cascade models and hydrodynamic simulations 
support this assumption).  
On the other hand, the question remains 
whether the dilepton yield from correlated open charm decays can be 
determined accurately enough to disentangle QGP signals. A combination  
of experimental (\eg, $e\mu$ coincidence measurements) and theoretical 
efforts (\eg, reliable estimates of energy loss effects in charm
quark propagation) might well achieve the required sensitivity.  

Finally, let us comment on further issues that are likely to
be addressed with potential upgrades of the PHENIX detector 
(such as additional vertex tracking devices~\cite{priv00}). 
First, a direct measurement of open charm production would improve  
the possibilities of studying direct QGP radiation substantially. 
Second, an evaluation of combinatorial background could facilitate 
a measurement of the low-mass continuum around the vector-meson
resonances, thus scrutinizing in-medium modifications of the $\rho$ 
meson in a net baryon-free environment. Since the isovector ($\rho$)
channel has a well-defined chiral partner in form of the $a_1$ 
channel, questions concerning  chiral symmetry restoration can be 
investigated. Similar connections are less
direct for $\omega$ or $\phi$ mesons.

\vskip1cm
 
\centerline {\bf ACKNOWLEDGMENTS}
I am  grateful for productive conversations with  A. Drees,
B. K\"ampfer, E.V. Shuryak, I. Tserruya and J. Wambach. 
This work was supported in part by the Alexander-von-Humboldt 
foundation (Feodor-Lynen program) and   
the U.S. Department of Energy under Grant No. DE-FG02-88ER40388.


%
%
%

\end{document}